\documentclass[a4paper,fleqn,usenatbib]{mnras}

\usepackage[T1]{fontenc}

\usepackage[english]{babel}
\usepackage{graphicx}
\usepackage{fleqn}
\usepackage{amssymb}
\usepackage{amsmath}
\usepackage{booktabs}
\usepackage{hyperref}
\usepackage{txfonts} 

\usepackage{color, colortbl}

\renewcommand{\S}{Section}
\newcommand{\F}{Fig.}

\newcommand{\ve}[1]{\boldsymbol{#1}}

\title[Secular evolution and HJs in multiplanet systems]{Secular dynamics of multiplanet systems: implications for the formation of hot and warm Jupiters via high-eccentricity migration}

\author[Hamers et al.]{Adrian S. Hamers$^{1}$\thanks{E-mail: hamers@ias.edu}, Fabio Antonini$^{2,3}$, Yoram Lithwick$^{2,3}$,  Hagai B. Perets$^{4}$ and \newauthor Simon F. Portegies Zwart$^{1}$ \\
$^{1}$Leiden Observatory, Leiden University, PO Box 9513, NL-2300 RA Leiden, The Netherlands \\
$^{2}$Center for Interdisciplinary Exploration and Research in Astrophysics (CIERA), Northwestern University, 2145 Sheridan Road, Evanston, IL 60208, USA \\
$^{3}$Department of Physics and Astronomy, Northwestern University, 2145 Sheridan Road, Evanston, IL 60208, USA \\
$^{4}$Technion - Israel Institute of Technology, Haifa 32000, Israel}
\date{Accepted 2016 September 15. Received 2016 September 12; in original form 2016 June 23}

\pubyear{2016}

\begin{document}
\label{firstpage}
\pagerange{\pageref{firstpage}--\pageref{lastpage}}
\maketitle

\begin{abstract} 
Hot Jupiters (HJs) are Jupiter-like planets that reside very closely to their host star, within $\sim 0.1\,\mathrm{AU}$. Their formation is not well understood. It is generally believed that they cannot have formed in situ, implying that some form of migration must have occurred after their initial formation. We study the production of HJs through secular evolution in multiplanet systems with three to five planets. In this variant of high-$e$ migration, the eccentricity of the orbit of the innermost planet is excited on secular time-scales, triggering orbital migration due to tidal dissipation. We use a secular dynamics code and carry out a population synthesis study. We find that HJs are only produced if the viscous time-scale is short ($\approx 0.014$ yr). In contrast, in up to $\approx 0.3$ of systems, the innermost planet is tidally disrupted. The orbital period distribution is peaked around 5 d, consistent with observations. The median HJ mass is $1\,M_\mathrm{J}$ with a maximum of $\approx 2 \, M_\mathrm{J}$, similar to observed HJs. Approximately 0.1 of the HJs have retrograde orbits with respect to the stellar spin. We do not find a significant population of warm Jupiters in our simulations, i.e. planets with semimajor axes between 0.1 and 1 AU. 
\end{abstract}


\section{Introduction}
\label{sect:introduction}
In the past decades, radial velocity and transit methods have revealed a population of gas giant planets of order Jupiter mass around solar-type stars with orbital periods downward of 10 d, i.e. hot Jupiters (HJs). The current consensus is that HJs could not have formed {\it in situ} in the protoplanetary disc phase because of an insufficient amount of disc material and/or too high temperatures at these close regions to the host star (e.g. \citealt{1996Natur.380..606L}; however, recently it has been suggested that {\it in situ} formation might be possible through core accretion, \citealt{2014ApJ...797...95L,2016ApJ...829..114B}). If HJs were formed at larger separations, i.e. beyond the snow line of one to a few AU, then this implies that they must have experienced strong inwards migration after their formation, by two orders of magnitude in separation. Two main migration scenarios have been proposed: (1) migration induced by orbital energy dissipation due to gas drag in the protoplanetary disc phase (e.g. \citealt{1980ApJ...241..425G,1986ApJ...309..846L,2000Icar..143....2B,2002ApJ...565.1257T}), and (2) migration induced by tidal dissipation in the HJ, requiring high orbital eccentricity (commonly known as `high-$e$' migration). 

For high-$e$ migration, various subscenarios have been proposed to drive the high eccentricities needed to produce the observed short orbital periods through tidal dissipation. They include (i) eccentricity excitation because of close encounters between planets \citep{1996Sci...274..954R,2008ApJ...686..580C,2008ApJ...686..621F,2008ApJ...686..603J,2012ApJ...751..119B}, (ii) excitation of the eccentricity because of secular Lidov-Kozai (LK) oscillations \citep{1962P&SS....9..719L,1962AJ.....67..591K} induced by a distant binary companion star or an additional (massive) planet on an inclined orbit \citep{2003ApJ...589..605W,2007ApJ...669.1298F,2012ApJ...754L..36N,2015ApJ...799...27P,2016MNRAS.456.3671A,2016ApJ...829..132P}, (iii) secular eccentricity excitation induced by a close and coplanar, but eccentric planetary companion \citep{2015ApJ...805...75P}, and (iv) eccentricity excitation induced by secular chaos in multiplanet systems with at least three planets in mildly inclined and eccentric orbits \citep{2011ApJ...735..109W,2011ApJ...739...31L,2014PNAS..11112610L}. 

It is currently unclear which of the two scenarios (1) and (2) applies to observed HJs, or if a combination gives rise to HJs. Both scenarios have successes and failures in describing properties of observed HJs. The observed period distribution of HJs peaks around $\sim 3-5 \, \mathrm{d}$, and their eccentricities are close to zero (e.g. \citealt{2016A&A...587A..64S}). Most high-$e$ migration scenarios predict that the final orbit of the HJ should indeed be circular and pile up a short period, around $\sim 3 \, \mathrm{d}$. Disc migration scenarios are more difficult to reconcile with the observed peak in the period distribution (but see \citealt{1996Natur.380..606L}). On the other hand, high-$e$ migration through LK cycles in three-body systems (subscenarios ii and iii) requires the presence of a stellar binary or compact planetary companion, which have not (yet) been detected around all HJs (e.g. \citealt{2014ApJ...785..126K,2015ApJ...800..138N}). Moreover, the predicted production rate is too low, and the predicted periods are too short (e.g. \citealt{2016MNRAS.456.3671A}).

Subscenario (iv) involves three or more planets around a single star, i.e. $N_\mathrm{p}\geq 3$ systems \citep{2011ApJ...735..109W,2011ApJ...739...31L,2014PNAS..11112610L}. Similarly to three-body systems (i.e. a binary companion or two planets), secular interactions can change the eccentricities of the orbits, in particular the innermost orbit, on long time-scales (i.e. much longer than the orbital periods). For $N_\mathrm{p}=2$, high relative inclinations (typically $\gtrsim 40^\circ$) and/or tight and eccentric outer orbits are required to produce high eccentricities. If $N_\mathrm{p}\geq 3$, then the conditions for producing high eccentricities in the innermost orbit are less stringent. The initial eccentricities and relative inclinations can be much smaller, and the planets need not be very closely spaced. Suborbital effects such as close encounters and mean motion resonances are then typically unimportant, and the long-term evolution is driven mainly by secular interactions. The secular evolution is typically chaotic, giving rise to highly irregular eccentricity oscillations. Over the course of $\sim 0.1-10 \,\mathrm{Gyr}$, high eccentricities can be reached in the inner orbit, potentially leading to HJs. A well known example is the Solar system, in which secular chaos can drive the orbit of Mercury to become unstable (i.e. lead to an ejection of the planet, or a collision of the planet with the Sun) on a time-scale of $5 \, \mathrm{Gyr}$, and with a probability of a few per cent \citep{2009Natur.459..817L,2014PNAS..11112610L}.

The less stringent conditions for secular chaos in $N_\mathrm{p} \geq 3$ systems are compatible with current observations of HJs which exclude close-in companions for a subset of HJs, whereas detections of further-away companions (at $\gtrsim 5-10 \, \mathrm{AU}$) are still largely incomplete because of observational limitations. This argument can also be reversed: the production of HJs through secular chaos {\it requires} further-away companions, therefore such companions are expected to be observed around HJ-hosting stars in the future. 

Parameter space studies and/or Monte Carlo studies to quantify observed properties of the HJs in multiplanet systems are still lacking. The long secular time-scales compared to the short orbital periods imply that direct $N$-body integrations, such as those carried out by \citet{2011ApJ...735..109W} and \citet{2012ApJ...751..119B}, are computationally very expensive to carry out, especially considering the large number of parameters. Until recently, secular, orbit-averaged methods valid for high eccentricities and inclinations were limited to $N_\mathrm{p}=2$ and to systems that are not too compact, as a consequence of the expansion of the Hamiltonian in terms of ratios of binary separations.

In recent work \citep{2016MNRAS.459.2827H}, we presented a generalization of the secular, orbit-averaged method previously applied to hierarchical three-body systems (e.g. \citealt{1962P&SS....9..719L,1962AJ.....67..591K,1968AJ.....73..190H,2013MNRAS.431.2155N}), to systems composed of nested binary orbits, with an arbitrary number of bodies and an arbitrary hierarchy, and to fifth order in terms of binary separation ratios for binary-binary interactions. In this paper, we apply this method to study the formation of HJs through secular evolution in multiplanet systems with $N_\mathrm{p}=3$ to $N_\mathrm{p}=5$ planets. The main practical advantage of this method is that compared to direct $N$-body integrations, the evolution can be computed much faster. 

Using a combination of population synthesis and grid sampling, we study the dependence of the HJ properties on various parameters, including the efficiency of tidal dissipation, the number of planets, the width of the initial mutual inclination and eccentricity distribution, and the radius of the innermost planet. These parameters, in particular the efficiency of tidal dissipation, are highly uncertain. 

The structure of this paper is as follows. In \S\,\ref{sect:methods}, we describe the secular method and other assumptions and in \S\,\ref{sect:ver}, we verify it by comparing to a (limited) number of direct $N$-body integrations. In \S\,\ref{sect:pop_syn}, we present the results from the population synthesis study. We discuss our results in \S\,\ref{sect:discussion}, and conclude in \S\,\ref{sect:conclusions}.

\section{Methods and assumptions}
\label{sect:methods}

\begin{table*}
\begin{tabular}{lp{4.5cm}cc}
\toprule
Symbol & Description & &  \\
\midrule
& & \multicolumn{2}{c}{Values in section} \\
& & \ref{sect:ver} & \ref{sect:pop_syn} \\
\cmidrule(l{2pt}r{2pt}){3-4} 
$N_\mathrm{p}$              & Number of planets             & 3                                                         & 3-5 \\
$M_\star$                   & Stellar mass                  & $1\,\mathrm{M}_\odot$                                     & $1\,\mathrm{M}_\odot$ \\
$m_i$                       & Mass of planet $i$            & $3\times 10^{-4} \, \mathrm{M}_\odot$ ($i=1$)             & 0.5-5 $M_\mathrm{J}$ \\
                            &                               & $1\times 10^{-4} \, \mathrm{M}_\odot$ ($i>1$)             & \\
$R_\star$                   & Stellar radius                & $1\,\mathrm{R}_\odot$                                     & $1\,\mathrm{R}_\odot$ \\
$R_i$                       & Radius of planet $i$          & $1\,R_\mathrm{J}$                                         & 1-1.5 $R_\mathrm{J}$ \\
$\eta$                      & Tidal disruption factor (cf. equation~\ref{eq:r_t}) & $-$                                 & 2.7 \\
$t_\mathrm{V,\star}$        & Stellar viscous time-scale    & $-$                                                       & $5 \, \mathrm{yr}$ \\
$t_\mathrm{V,1}$            & Planet 1 viscous time-scale   & $\approx 4.8\,\mathrm{yr}$                                &$0.0137,0.137,1.37\,\mathrm{yr}$ \\
$k_\mathrm{AM,\star}$       & Stellar apsidal motion constant & $-$                                                     & 0.014 \\
$k_\mathrm{AM,1}$           & Planet 1 apsidal motion constant & 0.19                                                   & 0.25 \\
$r_\mathrm{g,\star}$        & Stellar gyration radius       & $-$                                                       & 0.08 \\
$r_\mathrm{g,1}$            & Planet 1 gyration radius      & $-$                                                       & 0.25 \\
$P_\mathrm{s,\star}$        & Stellar spin period           & $-$                                                       & $10 \, \mathrm{d}$ \\
$P_\mathrm{s,1}$            & Planet 1 spin period          & $-$                                                       & $10 \, \mathrm{hr}$ \\
$\theta_\star$              & Stellar obliquity (stellar spin-orbit 1 angle) & $-$                                      & $0^\circ$ \\
$\theta_1$                  & Planet 1 obliquity (planet 1 spin-orbit 1 angle) & $-$                                    & $0^\circ$ \\
$a_i$                       & Planet $i$ orbital semimajor axis   & 1 AU ($i=1$)              & 1-4 AU ($i=1$) \\
                            &                               & 6 AU ($i=2$)                                              & 6-10 AU ($i=2$) \\
                            &                               & 12-62 AU ($i=3$)                                          & 15-30 AU ($i=3$) \\
                            &                               &                                                           & 35-50 AU ($i=4$) \\
                            &                               &                                                           & 60-100 AU ($i=5$) \\
$e_i$                       & Planet $i$ orbital eccentricity & $\approx$ 0.4-0.6                                       & 0-0.8 \\
$i_i$ & Planet $i$ orbital inclination\footnote{Defined with respect to an arbitrary inertial frame.}  &                & 0-30${}^\circ$ \\
$\omega_i$ & Planet $i$ argument of pericentre              & 0-360${}^\circ$                                           & 0-360${}^\circ$ \\
$\Omega_i$ & Planet $i$ longitude of ascending node         & 0-360${}^\circ$                                           & 0-360${}^\circ$ \\
$\beta$ & Width of inclination and eccentricity distribution & $-$ & 8.2, 14.6, 32.8 \\
\bottomrule
\end{tabular}
\caption{Description of the quantities used. Where applicable, we give the values of the (initial) parameters that are assumed in \S s\,\ref{sect:ver} and \ref{sect:pop_syn}. }
\label{table:IC}
\end{table*}

\subsection{Notations and overview}
In Table \ref{table:IC}, we give a list of relevant quantities with a description. Where applicable, we give for reference the values of the (initial) parameters that were assumed in the various sections of this paper.

\subsection{Secular dynamics}
\label{sect:methods:secular}
To model the long-term gravitational dynamics of the multiplanet system, we used the algorithm \textsc{SecularMultiple} \citep{2016MNRAS.459.2827H} within the \textsc{AMUSE} framework \citep{2013CoPhC.183..456P,2013A&A...557A..84P}. \textsc{SecularMultiple} applies to self-gravitating systems composed of nested binaries with an arbitrary number of bodies and an arbitrary hierarchy. A multiplanet system is represented as a `nested' hierarchical multiple system of point particles; the star is contained within the `innermost' binary system (i.e. Jacobian coordinates; see e.g. fig. 3 of \citealt{2016MNRAS.459.2827H}). The algorithm is based on an expansion of the Hamiltonian in terms of binary separation ratios, which are assumed to be small. The resulting Hamiltonian is orbit averaged, and the equations of motion, defining a system of ordinary differential equations, are solved numerically in terms of the orbital vectors $\ve{e}_k$ and $\ve{h}_k$ for all binaries $k$.

As shown in \citet{2016MNRAS.459.2827H}, depending on the compactness of the system, high orders are required to accurately describe the orbital evolution. Here, we included terms corresponding to binary-binary interactions (pairwise terms) up and including fifth order in the separation ratios. To third order (`octupole order'), we included the triplet binary terms (corresponding to interactions between three binaries), although these terms are unimportant for multiplanet systems with roughly equal-mass planets \citep{2016MNRAS.459.2827H}. For the fourth and fifth orders, terms associated with interactions between more than two binaries were not included; as shown in \citet{2016MNRAS.459.2827H}, these terms are unimportant compared to the pairwise binary terms.

Relativistic corrections were included by adding the orbit-averaged precession rates of the line of apsides to the orbits, to the first post-Newtonian (PN) order. Terms in the PN potential associated with interactions between binaries (e.g. \citealt{2013ApJ...773..187N}) were neglected (see also appendix A7 of \citealt{2016MNRAS.459.2827H}).

\subsection{Tidal evolution}
\label{sect:methods:tides}
The tidal evolution of the innermost planet and the star was modelled with the equilibrium tide model of \citet{1998ApJ...499..853E}. This model includes the effect of precession of the orbit of the innermost planet due to tidal bulges and rotation of both the star and innermost planet. We also included spin-orbit coupling and followed the spin directions of both the star and the innermost planet, assuming initially zero obliquities. The equilibrium tide model is described in terms of the viscous time-scale $t_\mathrm{V}$, the apsidal motion constant $k_\mathrm{AM}$, the gyration radius $r_\mathrm{g}$ and the initial spin period $P_\mathrm{s}$, for both the star and the innermost planet. Our assumed values are given in Table \ref{table:IC}. Most of the values are adopted from \citet{2007ApJ...669.1298F}.

We assumed a constant tidal viscous time-scale $t_\mathrm{V}$ for both the star and the innermost planet. Apart from its simplicity, a temporally constant $t_\mathrm{V,1}$ for the innermost planet during high-$e$ migration follows from the equations of motion with a number of physically motivated assumptions \citep{2012arXiv1209.5723S}. We note that our assumption of a constant $t_\mathrm{V,1}$ is in contrast to \citet{2003ApJ...589..605W,2011ApJ...735..109W}, who assumed a $t_\mathrm{V,1}$ depending on the orbital period \citep{2012arXiv1209.5724S}.

\section{Verification of the secular method with $N$-body integrations}
\label{sect:ver}
As mentioned in \S\,\ref{sect:introduction}, the secular evolution of multiplanet systems can be chaotic, especially when the number of planets is $N_\mathrm{p}\geq 3$. In direct $N$-body integrations, this implies that changing the initial orbital phases or the accuracy of the integration can lead to a completely different outcome after some time during which the eccentricities have changed by, say, the order of unity. This implies that it is not very meaningful -- on a one-to-one basis -- to compare results from direct $N$-body integrations to those from \textsc{SecularMultiple}, in which the orbits are averaged over. Our expectation is that \textsc{SecularMultiple} produces the correct secular dynamical evolution in a {\it statistical} sense, i.e. for an ensemble of systems. 

In this section, we investigate this expectation by comparing results from \textsc{SecularMultiple} to those from the direct $N$-body code \textsc{ARCHAIN} \citep{2008AJ....135.2398M}. The latter code uses algorithmic chain regularization to integrate the equations of motion with high precision. In addition to relativistic corrections, tidal interactions are taken into account with the same model assumed for the secular integrations. For more details regarding the direct $N$-body code, we refer to \citet{2016AJ....152..174A} and Antonini et al. (in preparation).

\subsection{Initial conditions}
\label{sect:ver:IC}
The \textsc{ARCHAIN} code was used to integrate 100 three-planet systems for $\approx 120 \, \mathrm{Myr}$ (cf. Table \ref{table:IC}). The stellar mass was set to $M_\star=1\,\mathrm{M}_\odot$, and the planetary masses were assumed to be $m_1 = 3\times 10^{-4} \, \mathrm{M}_\odot \approx 0.314 \, M_\mathrm{J}$ and $m_2=m_3 = 1\times10^{-3} \, \mathrm{M}_\odot \approx 1.05 \, M_\mathrm{J}$. The initial semimajor axes were assumed to be $a_1=1\,\mathrm{AU}$, $a_2=6\,\mathrm{AU}$, and $a_3$ was varied between 12 and 62 AU with increments of 0.5 AU. The eccentricities and inclinations for all orbits were set equal to each other, and values were assumed ranging between $\approx 0.4$ and $0.6$ (with the inclinations measured in radians). The other orbital angles, $\omega_i$ and $\Omega_i$, were sampled from flat distributions between 0 and 360${}^\circ$. The innermost planet was assumed to have a radius $R_1=1\,R_\mathrm{J}$, a viscous time-scale $t_\mathrm{V,1}\approx4.8\,\mathrm{yr}$, and an apsidal motion constant $k_\mathrm{AM,1} = 0.19$. Tidal disruptions were not checked for (in contrast to \S\,\ref{sect:pop_syn}).

With \textsc{ARCHAIN}, the typical integration time of a single system was $\sim$ 1 week (on a single CPU core). With comparable hardware, the corresponding integration time with \textsc{SecularMultiple} ranged between order 10 s (if there is no strong tidal evolution in the innermost orbit) to order a few min (if there is strong tidal evolution in the innermost orbit), corresponding to a speed-up factor between $\sim 10^3$ and $\sim 10^4$. 

\subsection{Results}
\label{sect:ver:results}
With \textsc{ARCHAIN}, a fraction of 0.32 of the systems became HJ systems by 120 Myr, i.e. the final semilatus rectum of the orbit of the innermost planet, $r_\mathrm{1,slr,f} \equiv a_{1,\mathrm{f}}\left(1-e_{1,\mathrm{f}}^2 \right )$, reached $r_\mathrm{1,slr,f} < 0.091 \, \mathrm{AU}$, corresponding to the semimajor axis of a 10-d planet in a circular orbit around a Solar-mass star. We use the semilatus rectum to define HJs, because after 120 Myr, some systems were still decaying tidally (i.e. $a_1$ and $e_1$ decreasing), while decoupled from the secular oscillations. In the latter case, when tidal evolution dominates, the final result (i.e. after $\gg 120 \, \mathrm{Myr}$) is a circular orbit with semimajor axis $a_\mathrm{1,f} = r_\mathrm{1,slr,f}$. The HJ fraction after 120 Myr of evolution obtained with \textsc{SecularMultiple} is 0.33.

\begin{figure}
\center
\includegraphics[scale = 0.45, trim = 10mm 0mm 0mm 0mm]{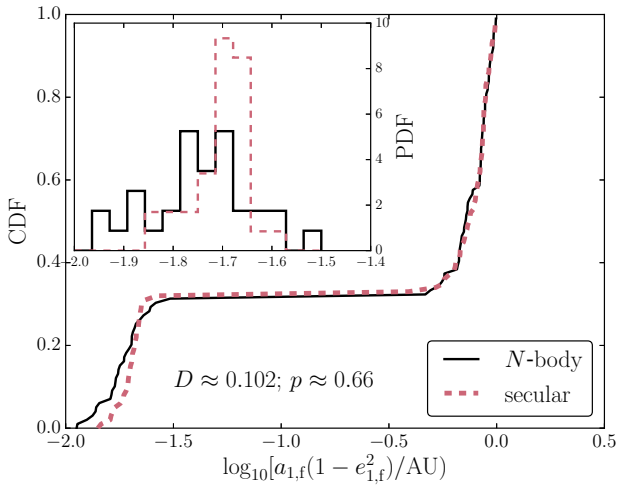}
\caption{\small The distributions of the innermost planet's orbit semilatus rectum, $r_\mathrm{1,slr,f} \equiv a_{1,\mathrm{f}}\left(1-e_{1,\mathrm{f}}^2 \right )$, after $120 \, \mathrm{Myr}$ of evolution according to \textsc{ARCHAIN} (black solid line) and to \textsc{SecularMultiple} (red dashed line). The $D$ and $p$ statistics for the two-sided KS test between the distributions are indicated in the panel. In the inset, the PDF is shown for $\log_{10} (r_\mathrm{1,slr,f}/\mathrm{AU})<-1.5$. }
\label{fig:nbody_ref_14_03_16_120Myr_d_tV0_final_sma_distribution}
\end{figure}

In \F\,\ref{fig:nbody_ref_14_03_16_120Myr_d_tV0_final_sma_distribution}, we show the distribution of $r_\mathrm{1,slr,f}$ after $120 \, \mathrm{Myr}$ of evolution in terms of the cumulative density function (CDF) according to \textsc{ARCHAIN} (black solid line) and according to \textsc{SecularMultiple} (red dashed line). There is a pileup of systems around $\log_{10}(r_\mathrm{1,slr,f}/\mathrm{AU}) \approx -1.7$ corresponding to systems in which a HJ was formed. The number of systems subsequently stalls at $\approx 0.32$ with increasing $r_\mathrm{1,slr,f}$, until at $\log_{10}(r_\mathrm{1,slr,f}/\mathrm{AU}) \approx -0.4$, this number increases again. The latter correspond to non- (or weakly) migrating planets; either their semimajor axes have decreased slightly due to tidal evolution, and/or their eccentricities are excited because of secular evolution. 

In terms of $r_\mathrm{1,slr,f}$, there is statistical agreement between \textsc{ARCHAIN} and \textsc{SecularMultiple}; the two-sided Kolmogorov-Smirnov (KS; \citealt{kolmogorov_33,smirnov_48}) test statistics are $D\approx 0.082$ and $p\approx 0.881$. There is a tendency for \textsc{SecularMultiple} to (slightly) overestimate the smallest semilatus recta, or the smallest orbital periods of the HJs, compared to \textsc{ARCHAIN}. We do not believe that this discrepancy strongly affects our conclusions in the integrations in \S\,\ref{sect:pop_syn}. 

\begin{figure}
\center
\includegraphics[scale = 0.45, trim = 10mm 0mm 0mm 0mm]{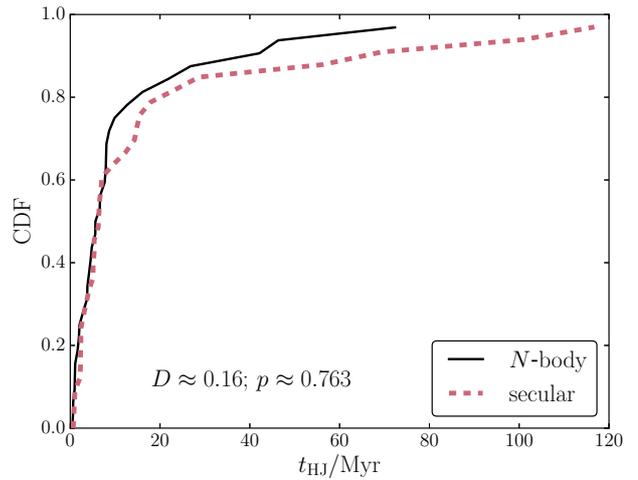}
\caption{\small The CDF of the HJ formation times (see text for definition) according to \textsc{ARCHAIN} (black solid line) and to \textsc{SecularMultiple} (red dashed line). KS test statistics are indicated. }
\label{fig:nbody_ref_14_03_16_120Myr_d_tV0_HJ_formation_times}
\end{figure}

In \F\,\ref{fig:nbody_ref_14_03_16_120Myr_d_tV0_HJ_formation_times}, we show another comparison between the two methods in terms of the distributions of the HJ formation times, i.e. the times when $r_\mathrm{1,slr,f}\lesssim 0.09\,\mathrm{AU}$ (corresponding to an orbital period of 10 d around a solar-mass star). With both methods, most ($\approx 0.7$) HJs are produced early in the evolution, i.e. within the first 10 Myr. The KS test yields $D\approx 0.16$ and $p\approx0.763$, showing that the two distributions are statistically consistent. Nonetheless, the formation times in \textsc{SecularMultiple} after 10 Myr tend to be slightly longer. 

Despite the statistical consistencies as described above, it should be taken into account that the condition $K_{ij}\leq 0$ occurred in both the secular and direct integrations. The quantity $K_{ij}$ is defined for two adjacent orbits $i$ and $j$, with $a_j>a_i$, as
\begin{align}
\label{eq:K_def}
K_{ij} \equiv \frac{ a_j(1-e_j) - a_i(1+e_i)}{R_{\mathrm{H};ij}},
\end{align}
where
\begin{align}
R_{\mathrm{H};ij} \equiv \frac{a_i+a_j}{2} \left ( \frac{m_i+m_j}{3M_\star} \right )^{1/3}
\end{align}
is the mutual Hill radius. With \textsc{SecularMultiple}, $K_{ij}\leq0$ occurred for $\approx 0.00016$ and $\approx 0.021$ of all output times for orbit pairs $(1,2)$ and $(2,3)$, respectively. With \textsc{ARCHAIN}, these fractions were $\approx 0.0002$ and $\approx 0.015$, for the same orbit pairs. If $K_{ij}\leq0$, then the assumptions on which the secular method is based formally break down, and results should be interpreted critically. In our population synthesis study (\S\,\ref{sect:pop_syn}), integrations were stopped at occurrences of $K_{ij}\leq0$. To investigate whether this affects our conclusions, some of the integrations were not stopped when $K_{ij}\leq0$.

\section{Population synthesis}
\label{sect:pop_syn}
We used \textsc{SecularMultiple} to compute the dynamical evolution of a population of multiplanet systems with three to five planets. The innermost planet was initially located between 1 and 4 AU commensurate with planet formation just beyond the snow line. Taking into account tidal evolution, we focus on the orbital evolution of the innermost planet in the context of WJs and HJs.

\subsection{Initial conditions}
\label{sect:pop_syn:IC}
To generate the initial conditions of our systems, we used a combination of grid and Monte Carlo sampling. We defined a grid with the following parameters: (1) the viscous time-scale $t_\mathrm{V,1}$ of the innermost planet, (2) the number of planets $N_\mathrm{p}$, (3) $\beta$, which is related to the rms width of the initial inclination and eccentricity distribution (see below), and (4) the radius of the innermost planet, $R_1$. The following values were considered:
\begin{align}
\left \{
\begin{array}{ll}
t_\mathrm{V,1} &\in \left \{10^{-3},10^{-2},10^{-1},10^0 \right \} \, t_\mathrm{V,SKD}; \\
N_\mathrm{p} &\in \{3,4,5\}; \\
\beta &\in \left \{ \left (20 \frac{\pi}{180} \right )^{-2}, \left (15 \frac{\pi}{180} \right )^{-2},\left (10 \frac{\pi}{180} \right )^{-2} \right \} \\
& \approx \{8.2, 14.6, 32.8 \} ; \\
R_1 &\in \{1.0,1.5\} \, R_\mathrm{J}.
\end{array} \right.
\end{align}
For gas giant planets and high-$e$ migration, \citet{2012arXiv1209.5724S} provided the constraint $t_\mathrm{V,1} \lesssim 1.2 \, \times 10^{4} \, \mathrm{hr}$, by requiring that a HJ at 5 d is circularized in less than 10 Gyr. We adopted a reference value $t_\mathrm{V,SKD} = 1.2 \times 10^4 \, \mathrm{h} \approx 1.37 \, \mathrm{yr}$ , and considered $t_\mathrm{V,1}$ corresponding to 1, 10, 100 and 1000 times more efficient tides compared to $t_\mathrm{V,SKD}$ (smaller $t_\mathrm{V,1}$ correspond to more efficient tides). The viscous time-scale $t_\mathrm{V,SKD}$ corresponds to a tidal quality factor of $Q_1\approx 1.1\times 10^5$ (cf. equation 37 from \citealt{2012arXiv1209.5724S}), and $Q_1 \propto t_\mathrm{V,1}$.

We considered two planetary radii of 1 and 1.5 $R_\mathrm{J}$. Large planetary radii might be expected because of inflation due to tidal heating, and are also observed (this is known as the radius anomaly, \citealt{2011ApJ...729L...7L}). 

For each of the 72 combinations of grid parameters, we sampled $N_\mathrm{MC}=1000$ systems using the following approach. The masses $m_i$ of the $N_\mathrm{p}$ planets were sampled from flat distributions with $0.5 \, M_\mathrm{J} < m_i < 5 \, M_\mathrm{J}$. The inclinations $i_i$ (as measured in radians, and defined with respect to an arbitrary reference plane) and eccentricities $e_i$ of the planetary orbits were both sampled from a Rayleigh distribution,
\begin{align}
\frac{\mathrm{d} N}{\mathrm{d} x} \propto x \exp \left (-\beta x^2 \right ),
\end{align}
where $\beta = \langle x^2 \rangle^{-1}$ (assuming $0<x<\infty$) characterizes the width of the distribution. The sampling limits were $0<e_i<0.8$ and $0^\circ<i_i < 30^\circ$. The arguments of pericentre $\omega_i$ and the longitudes of the ascending nodes $\Omega_i$ were sampled from flat distributions between $0^\circ$ and $360^\circ$.

The semimajor axes of the planets $a_i$ were sampled from flat distributions, with the fixed ranges
\begin{align}
\left \{
\begin{array}{llll}
1.0 \, \mathrm{AU} &< &a_1 &< 4.0 \, \mathrm{AU}; \\
6.0 \, \mathrm{AU} &< &a_2 &< 10.0 \, \mathrm{AU}; \\
15.0 \, \mathrm{AU} &< &a_3 &< 30.0 \, \mathrm{AU}; \\
35.0 \, \mathrm{AU} &< &a_4 &< 50.0 \, \mathrm{AU}; \\
60.0 \, \mathrm{AU} &< &a_5 &< 100.0 \, \mathrm{AU}.
\end{array} \right.
\end{align}
These choices are somewhat arbitrary. The semimajor axes of planets beyond $\sim 5 \, \mathrm{AU}$ are still poorly constrained by observations. Assumptions must therefore be made regarding the semimajor axes, in particular for orbits outside of the innermost planet. The ranges of $a_1$, $a_2$ and $a_3$ are similar to the values that were assumed by \citet{2011ApJ...735..109W} (1, 6 and 16 AU, respectively). The choice of a flat distribution, apart from its simplicity, is motivated by the ability to easily disentangle any dependence of the results on the semimajor axes. At any rate, it should be taken into account that our choice of the initial semimajor axes is not unique, and likely affects the results of the population synthesis. 

We rejected a sampled combination of $m_i$, $a_i$ and $e_i$ if $K_{ij}<K_0\equiv 2$ for any adjacent pair $(i,j)=(i,i+1)$ of orbits (cf. equation~\ref{eq:K_def}). The specific value of $K_0$ is arbitrary. Choosing the value $K_0=0$ would produce a large fraction of systems in which the secular method is questionable from the start (cf. \S\,\ref{sect:ver}), whereas a large value of $K_0$, say $K_0=12$, would produce too-well-separated systems in which the eccentricities would hardly evolve, and no WJs or HJs would be formed. As a compromise, we set $K_0=2$.

For the other (non-sampled) initial parameters, we refer to the last column of Table\,\ref{table:IC}. Regarding the star, we adopted a constant viscous time-scale of $t_\mathrm{V,\star} = 5 \, \mathrm{yr}$. Assuming a (massless) companion at an orbital period of 4 d, this corresponds to a tidal quality factor of $Q \sim 6 \times 10^8$ or $Q' \equiv 3Q/(4 k_\mathrm{AM,\star}) \sim 3 \times 10^6$, which is typical for Solar-type stars \citep{2007ApJ...661.1180O}.

\subsection{Stopping conditions}
\label{sect:pop_syn:SC}
The integrations were stopped if one of the following conditions was met.
\begin{enumerate}
\item The integration time reached 10 Gyr.
\item A HJ was formed, i.e. $P_1 < 10 \, \mathrm{d}$ and $e_1 < 10^{-3}$. The orbit is then well within the regime of the decoupling of tidal from secular evolution, and there is no further evolution due to tides raised on the planet. Tides raised on the star are still important, but the time-scale for inspiral, $\sim 20 \, \mathrm{Gyr}$ (assuming a stellar viscous time-scale of 5 yr), is longer than the Hubble time.
\item The innermost planet was tidally disrupted by the star, i.e. $r_\mathrm{p,1}=a_1(1-e_1)<r_\mathrm{t}$, where $r_\mathrm{t}$ is given by
\begin{align}
\label{eq:r_t}
r_\mathrm{t} = \eta R_1 \left (\frac{ M_\star }{m_1} \right )^{1/3}.
\end{align}
Here, $\eta$ is a dimensionless parameter; throughout, we assumed $\eta=2.7$ \citep{2011ApJ...732...74G}. 
\item The condition $K_{ij}\leq0$ (cf. equation~\ref{eq:K_def}) occurred for any pair of adjacent orbits $(i,j)$, i.e. the secular approximation formally broke down.  We also carried out additional simulations for some parameter combinations without this stopping condition to investigate the sensitivity of our results on this  condition.
\item The run time of the simulation exceeded 12 CPU hours (imposed for practical reasons).\footnote{Depending on the parameters, the secular time-scales can be very short compared to the integration time of 10 Gyr, implying that a very large number of oscillations need to be computed. Consequently, some simulations can take several hours to complete, in contrast to \S\,\ref{sect:ver}, in which the secular time-scales are typically short compared to the integration time of 120 Myr (see also \S\,\ref{sect:discussion:uncertainties}). }
\end{enumerate}

\definecolor{Gray}{gray}{0.9}
\begin{table*}
\scriptsize
\begin{tabular}{cccccccccccccccccccc}
\toprule
& & & & \multicolumn{3}{c}{$f_\mathrm{no\, migration}$} & \multicolumn{3}{c}{$f_\mathrm{HJ}$} & \multicolumn{3}{c}{$f_\mathrm{TD}$} & \multicolumn{3}{c}{$f_{K_{ij}\leq0}$} & \multicolumn{3}{c}{$f_\mathrm{run\, time\,exceeded}$} \\
\cmidrule(l{2pt}r{2pt}){5-7} \cmidrule(l{2pt}r{2pt}){8-10} \cmidrule(l{2pt}r{2pt}){11-13}  \cmidrule(l{2pt}r{2pt}){14-16} \cmidrule(l{2pt}r{2pt}){17-19}
& & & & \multicolumn{2}{c}{$t_\mathrm{end} / \mathrm{Gyr}$} & & \multicolumn{2}{c}{$t_\mathrm{end} / \mathrm{Gyr}$} & & \multicolumn{2}{c}{$t_\mathrm{end} /\mathrm{Gyr}$} & & \multicolumn{2}{c}{$t_\mathrm{end} / \mathrm{Gyr}$} & & \multicolumn{2}{c}{$t_\mathrm{end} / \mathrm{Gyr}$} & \\
\cmidrule(l{2pt}r{2pt}){5-6} \cmidrule(l{2pt}r{2pt}){8-9} \cmidrule(l{2pt}r{2pt}){11-12}  \cmidrule(l{2pt}r{2pt}){14-15} \cmidrule(l{2pt}r{2pt}){17-18}
$t_\mathrm{V,1}/\mathrm{yr}$ & $N_\mathrm{p}$ & $\beta$ & $R_1/R_\mathrm{J}$ & 5 & 10 & $t_x$ & 5 & 10 & $t_x$ & 5 & 10 & $t_x$ & 5 & 10 & $t_x$ & 5 & 10 & $t_x$ \\
\midrule
$1.37 \times 10^{-3}$ & 3 & 8.2 & 1.0 & 0.822 & 0.816 & 0.831 & 0.008 & 0.008 & 0.007 & 0.048 & 0.049 & 0.047 & 0.122 & 0.127 & 0.115 & 0.000 & 0.000 & 0.000 &  \\
\rowcolor{Gray}
$1.37 \times 10^{-3}$ & 3 & 8.2 & 1.5 & 0.820 & 0.810 & 0.828 & 0.021 & 0.023 & 0.018 & 0.020 & 0.020 & 0.019 & 0.138 & 0.146 & 0.134 & 0.000 & 0.000 & 0.000 &  \\
$1.37 \times 10^{-3}$ & 3 & 14.6 & 1.0 & 0.912 & 0.906 & 0.916 & 0.010 & 0.010 & 0.009 & 0.022 & 0.023 & 0.021 & 0.056 & 0.061 & 0.054 & 0.000 & 0.000 & 0.000 &  \\
\rowcolor{Gray}
$1.37 \times 10^{-3}$ & 3 & 14.6 & 1.5 & 0.929 & 0.921 & 0.933 & 0.011 & 0.012 & 0.010 & 0.010 & 0.011 & 0.010 & 0.050 & 0.054 & 0.045 & 0.000 & 0.000 & 0.000 &  \\
$1.37 \times 10^{-3}$ & 3 & 32.8 & 1.0 & 0.987 & 0.986 & 0.987 & 0.000 & 0.000 & 0.000 & 0.003 & 0.003 & 0.003 & 0.010 & 0.011 & 0.010 & 0.000 & 0.000 & 0.000 &  \\
\rowcolor{Gray}
$1.37 \times 10^{-3}$ & 3 & 32.8 & 1.5 & 0.990 & 0.989 & 0.990 & 0.001 & 0.001 & 0.001 & 0.001 & 0.001 & 0.001 & 0.008 & 0.009 & 0.008 & 0.000 & 0.000 & 0.000 &  \\
$1.37 \times 10^{-3}$ & 4 & 8.2 & 1.0 & 0.347 & 0.321 & 0.372 & 0.007 & 0.009 & 0.006 & 0.055 & 0.058 & 0.056 & 0.591 & 0.611 & 0.566 & 0.000 & 0.001 & 0.000 &  \\
\rowcolor{Gray}
$1.37 \times 10^{-3}$ & 4 & 8.2 & 1.5 & 0.324 & 0.298 & 0.362 & 0.014 & 0.018 & 0.013 & 0.062 & 0.062 & 0.059 & 0.600 & 0.622 & 0.566 & 0.000 & 0.000 & 0.000 &  \\
$1.37 \times 10^{-3}$ & 4 & 14.6 & 1.0 & 0.580 & 0.547 & 0.591 & 0.006 & 0.006 & 0.006 & 0.043 & 0.045 & 0.041 & 0.371 & 0.401 & 0.362 & 0.000 & 0.001 & 0.000 &  \\
\rowcolor{Gray}
$1.37 \times 10^{-3}$ & 4 & 14.6 & 1.5 & 0.552 & 0.520 & 0.579 & 0.015 & 0.015 & 0.015 & 0.032 & 0.034 & 0.031 & 0.400 & 0.429 & 0.374 & 0.000 & 0.000 & 0.000 &  \\
$1.37 \times 10^{-3}$ & 4 & 32.8 & 1.0 & 0.867 & 0.858 & 0.881 & 0.002 & 0.002 & 0.002 & 0.010 & 0.010 & 0.008 & 0.121 & 0.130 & 0.109 & 0.000 & 0.000 & 0.000 &  \\
\rowcolor{Gray}
$1.37 \times 10^{-3}$ & 4 & 32.8 & 1.5 & 0.886 & 0.870 & 0.887 & 0.007 & 0.007 & 0.005 & 0.003 & 0.003 & 0.003 & 0.102 & 0.117 & 0.103 & 0.000 & 0.001 & 0.000 &  \\
$1.37 \times 10^{-3}$ & 5 & 8.2 & 1.0 & 0.117 & 0.092 & 0.139 & 0.005 & 0.006 & 0.005 & 0.045 & 0.045 & 0.043 & 0.833 & 0.853 & 0.811 & 0.000 & 0.004 & 0.002 &  \\
\rowcolor{Gray}
$1.37 \times 10^{-3}$ & 5 & 8.2 & 1.5 & 0.107 & 0.085 & 0.130 & 0.008 & 0.008 & 0.007 & 0.057 & 0.057 & 0.054 & 0.828 & 0.844 & 0.809 & 0.000 & 0.006 & 0.000 &  \\
$1.37 \times 10^{-3}$ & 5 & 14.6 & 1.0 & 0.239 & 0.209 & 0.275 & 0.005 & 0.005 & 0.005 & 0.051 & 0.051 & 0.051 & 0.705 & 0.732 & 0.669 & 0.000 & 0.003 & 0.000 &  \\
\rowcolor{Gray}
$1.37 \times 10^{-3}$ & 5 & 14.6 & 1.5 & 0.264 & 0.214 & 0.289 & 0.015 & 0.018 & 0.016 & 0.045 & 0.045 & 0.043 & 0.676 & 0.708 & 0.649 & 0.000 & 0.014 & 0.002 &  \\
$1.37 \times 10^{-3}$ & 5 & 32.8 & 1.0 & 0.669 & 0.612 & 0.689 & 0.005 & 0.005 & 0.005 & 0.025 & 0.026 & 0.022 & 0.301 & 0.325 & 0.279 & 0.000 & 0.032 & 0.005 &  \\
\rowcolor{Gray}
$1.37 \times 10^{-3}$ & 5 & 32.8 & 1.5 & 0.660 & 0.592 & 0.673 & 0.006 & 0.006 & 0.005 & 0.019 & 0.021 & 0.020 & 0.312 & 0.362 & 0.299 & 0.000 & 0.016 & 0.000 &  \\
\midrule
$1.37 \times 10^{-2}$ & 3 & 8.2 & 1.0 & 0.835 & 0.818 & 0.846 & 0.002 & 0.004 & 0.002 & 0.028 & 0.032 & 0.027 & 0.135 & 0.146 & 0.125 & 0.000 & 0.000 & 0.000 &  \\
\rowcolor{Gray}
$1.37 \times 10^{-2}$ & 3 & 8.2 & 1.5 & 0.819 & 0.807 & 0.830 & 0.002 & 0.005 & 0.001 & 0.046 & 0.048 & 0.045 & 0.133 & 0.140 & 0.124 & 0.000 & 0.000 & 0.000 &  \\
$1.37 \times 10^{-2}$ & 3 & 14.6 & 1.0 & 0.913 & 0.908 & 0.916 & 0.001 & 0.001 & 0.000 & 0.021 & 0.022 & 0.021 & 0.065 & 0.069 & 0.063 & 0.000 & 0.000 & 0.000 &  \\
\rowcolor{Gray}
$1.37 \times 10^{-2}$ & 3 & 14.6 & 1.5 & 0.920 & 0.912 & 0.924 & 0.001 & 0.001 & 0.001 & 0.020 & 0.020 & 0.018 & 0.059 & 0.067 & 0.057 & 0.000 & 0.000 & 0.000 &  \\
$1.37 \times 10^{-2}$ & 3 & 32.8 & 1.0 & 0.990 & 0.990 & 0.991 & 0.001 & 0.001 & 0.001 & 0.003 & 0.003 & 0.003 & 0.006 & 0.006 & 0.005 & 0.000 & 0.000 & 0.000 &  \\
\rowcolor{Gray}
$1.37 \times 10^{-2}$ & 3 & 32.8 & 1.5 & 0.988 & 0.988 & 0.988 & 0.001 & 0.001 & 0.001 & 0.003 & 0.003 & 0.003 & 0.008 & 0.008 & 0.008 & 0.000 & 0.000 & 0.000 &  \\
$1.37 \times 10^{-2}$ & 4 & 8.2 & 1.0 & 0.305 & 0.289 & 0.335 & 0.000 & 0.000 & 0.000 & 0.068 & 0.071 & 0.068 & 0.627 & 0.638 & 0.596 & 0.000 & 0.002 & 0.001 &  \\
\rowcolor{Gray}
$1.37 \times 10^{-2}$ & 4 & 8.2 & 1.5 & 0.334 & 0.300 & 0.343 & 0.002 & 0.002 & 0.002 & 0.063 & 0.063 & 0.062 & 0.601 & 0.635 & 0.593 & 0.000 & 0.000 & 0.000 &  \\
$1.37 \times 10^{-2}$ & 4 & 14.6 & 1.0 & 0.565 & 0.544 & 0.582 & 0.000 & 0.000 & 0.000 & 0.052 & 0.052 & 0.050 & 0.383 & 0.404 & 0.368 & 0.000 & 0.000 & 0.000 &  \\
\rowcolor{Gray}
$1.37 \times 10^{-2}$ & 4 & 14.6 & 1.5 & 0.565 & 0.531 & 0.584 & 0.002 & 0.003 & 0.003 & 0.037 & 0.038 & 0.035 & 0.396 & 0.426 & 0.378 & 0.000 & 0.002 & 0.000 &  \\
$1.37 \times 10^{-2}$ & 4 & 32.8 & 1.0 & 0.887 & 0.872 & 0.897 & 0.000 & 0.000 & 0.000 & 0.008 & 0.008 & 0.007 & 0.105 & 0.118 & 0.096 & 0.000 & 0.002 & 0.000 &  \\
\rowcolor{Gray}
$1.37 \times 10^{-2}$ & 4 & 32.8 & 1.5 & 0.884 & 0.866 & 0.883 & 0.003 & 0.003 & 0.002 & 0.005 & 0.006 & 0.005 & 0.108 & 0.121 & 0.108 & 0.000 & 0.004 & 0.002 &  \\
$1.37 \times 10^{-2}$ & 5 & 8.2 & 1.0 & 0.086 & 0.068 & 0.112 & 0.000 & 0.000 & 0.000 & 0.050 & 0.051 & 0.050 & 0.864 & 0.878 & 0.838 & 0.000 & 0.003 & 0.000 &  \\
\rowcolor{Gray}
$1.37 \times 10^{-2}$ & 5 & 8.2 & 1.5 & 0.106 & 0.091 & 0.128 & 0.002 & 0.002 & 0.001 & 0.068 & 0.069 & 0.066 & 0.824 & 0.831 & 0.804 & 0.000 & 0.007 & 0.001 &  \\
$1.37 \times 10^{-2}$ & 5 & 14.6 & 1.0 & 0.243 & 0.201 & 0.266 & 0.001 & 0.001 & 0.001 & 0.059 & 0.059 & 0.057 & 0.696 & 0.729 & 0.675 & 0.001 & 0.010 & 0.001 &  \\
\rowcolor{Gray}
$1.37 \times 10^{-2}$ & 5 & 14.6 & 1.5 & 0.246 & 0.202 & 0.274 & 0.001 & 0.001 & 0.001 & 0.052 & 0.054 & 0.051 & 0.701 & 0.728 & 0.670 & 0.000 & 0.015 & 0.004 &  \\
$1.37 \times 10^{-2}$ & 5 & 32.8 & 1.0 & 0.684 & 0.633 & 0.689 & 0.000 & 0.000 & 0.000 & 0.025 & 0.027 & 0.023 & 0.290 & 0.322 & 0.283 & 0.001 & 0.018 & 0.005 &  \\
\rowcolor{Gray}
$1.37 \times 10^{-2}$ & 5 & 32.8 & 1.5 & 0.640 & 0.586 & 0.653 & 0.001 & 0.001 & 0.001 & 0.023 & 0.024 & 0.019 & 0.336 & 0.369 & 0.319 & 0.000 & 0.020 & 0.008 &  \\
\midrule
$1.37 \times 10^{-1}$ & 3 & 8.2 & 1.0 & 0.807 & 0.798 & 0.810 & 0.000 & 0.000 & 0.000 & 0.041 & 0.042 & 0.041 & 0.152 & 0.160 & 0.149 & 0.000 & 0.000 & 0.000 &  \\
\rowcolor{Gray}
$1.37 \times 10^{-1}$ & 3 & 8.2 & 1.5 & 0.815 & 0.804 & 0.827 & 0.000 & 0.000 & 0.000 & 0.038 & 0.040 & 0.035 & 0.147 & 0.156 & 0.138 & 0.000 & 0.000 & 0.000 &  \\
$1.37 \times 10^{-1}$ & 3 & 14.6 & 1.0 & 0.921 & 0.913 & 0.922 & 0.000 & 0.000 & 0.000 & 0.023 & 0.023 & 0.021 & 0.056 & 0.064 & 0.057 & 0.000 & 0.000 & 0.000 &  \\
\rowcolor{Gray}
$1.37 \times 10^{-1}$ & 3 & 14.6 & 1.5 & 0.924 & 0.921 & 0.928 & 0.000 & 0.000 & 0.000 & 0.022 & 0.022 & 0.022 & 0.054 & 0.057 & 0.050 & 0.000 & 0.000 & 0.000 &  \\
$1.37 \times 10^{-1}$ & 3 & 32.8 & 1.0 & 0.990 & 0.989 & 0.991 & 0.000 & 0.000 & 0.000 & 0.003 & 0.003 & 0.003 & 0.007 & 0.008 & 0.006 & 0.000 & 0.000 & 0.000 &  \\
\rowcolor{Gray}
$1.37 \times 10^{-1}$ & 3 & 32.8 & 1.5 & 0.988 & 0.988 & 0.988 & 0.000 & 0.000 & 0.000 & 0.000 & 0.000 & 0.000 & 0.012 & 0.012 & 0.012 & 0.000 & 0.000 & 0.000 &  \\
$1.37 \times 10^{-1}$ & 4 & 8.2 & 1.0 & 0.355 & 0.331 & 0.377 & 0.000 & 0.000 & 0.000 & 0.059 & 0.060 & 0.057 & 0.586 & 0.607 & 0.566 & 0.000 & 0.002 & 0.000 &  \\
\rowcolor{Gray}
$1.37 \times 10^{-1}$ & 4 & 8.2 & 1.5 & 0.376 & 0.348 & 0.391 & 0.000 & 0.000 & 0.000 & 0.052 & 0.052 & 0.051 & 0.572 & 0.599 & 0.558 & 0.000 & 0.001 & 0.000 &  \\
$1.37 \times 10^{-1}$ & 4 & 14.6 & 1.0 & 0.536 & 0.499 & 0.557 & 0.000 & 0.000 & 0.000 & 0.046 & 0.046 & 0.046 & 0.418 & 0.453 & 0.397 & 0.000 & 0.002 & 0.000 &  \\
\rowcolor{Gray}
$1.37 \times 10^{-1}$ & 4 & 14.6 & 1.5 & 0.551 & 0.515 & 0.574 & 0.000 & 0.000 & 0.000 & 0.035 & 0.036 & 0.033 & 0.414 & 0.442 & 0.392 & 0.000 & 0.007 & 0.001 &  \\
$1.37 \times 10^{-1}$ & 4 & 32.8 & 1.0 & 0.893 & 0.872 & 0.898 & 0.000 & 0.001 & 0.001 & 0.008 & 0.009 & 0.007 & 0.099 & 0.117 & 0.094 & 0.000 & 0.001 & 0.000 &  \\
\rowcolor{Gray}
$1.37 \times 10^{-1}$ & 4 & 32.8 & 1.5 & 0.885 & 0.868 & 0.889 & 0.000 & 0.000 & 0.000 & 0.016 & 0.017 & 0.012 & 0.099 & 0.114 & 0.099 & 0.000 & 0.001 & 0.000 &  \\
$1.37 \times 10^{-1}$ & 5 & 8.2 & 1.0 & 0.111 & 0.081 & 0.128 & 0.000 & 0.000 & 0.000 & 0.066 & 0.067 & 0.066 & 0.823 & 0.843 & 0.805 & 0.000 & 0.009 & 0.001 &  \\
\rowcolor{Gray}
$1.37 \times 10^{-1}$ & 5 & 8.2 & 1.5 & 0.123 & 0.095 & 0.136 & 0.000 & 0.000 & 0.000 & 0.055 & 0.056 & 0.054 & 0.822 & 0.845 & 0.809 & 0.000 & 0.004 & 0.001 &  \\
$1.37 \times 10^{-1}$ & 5 & 14.6 & 1.0 & 0.262 & 0.220 & 0.286 & 0.000 & 0.000 & 0.000 & 0.052 & 0.053 & 0.049 & 0.685 & 0.719 & 0.663 & 0.001 & 0.008 & 0.002 &  \\
\rowcolor{Gray}
$1.37 \times 10^{-1}$ & 5 & 14.6 & 1.5 & 0.261 & 0.227 & 0.293 & 0.000 & 0.000 & 0.000 & 0.047 & 0.050 & 0.045 & 0.692 & 0.718 & 0.661 & 0.000 & 0.005 & 0.001 &  \\
$1.37 \times 10^{-1}$ & 5 & 32.8 & 1.0 & 0.688 & 0.628 & 0.703 & 0.000 & 0.000 & 0.000 & 0.033 & 0.040 & 0.033 & 0.279 & 0.314 & 0.259 & 0.000 & 0.018 & 0.005 &  \\
\rowcolor{Gray}
$1.37 \times 10^{-1}$ & 5 & 32.8 & 1.5 & 0.666 & 0.615 & 0.690 & 0.000 & 0.000 & 0.000 & 0.031 & 0.033 & 0.032 & 0.303 & 0.339 & 0.277 & 0.000 & 0.013 & 0.001 &  \\
\midrule
$1.37 \times 10^{0}$ & 3 & 8.2 & 1.0 & 0.835 & 0.824 & 0.845 & 0.000 & 0.000 & 0.000 & 0.035 & 0.039 & 0.033 & 0.130 & 0.137 & 0.122 & 0.000 & 0.000 & 0.000 &  \\
\rowcolor{Gray}
$1.37 \times 10^{0}$ & 3 & 8.2 & 1.5 & 0.844 & 0.836 & 0.856 & 0.000 & 0.000 & 0.000 & 0.038 & 0.039 & 0.037 & 0.118 & 0.125 & 0.107 & 0.000 & 0.000 & 0.000 &  \\
$1.37 \times 10^{0}$ & 3 & 14.6 & 1.0 & 0.916 & 0.908 & 0.916 & 0.000 & 0.000 & 0.000 & 0.015 & 0.016 & 0.014 & 0.069 & 0.076 & 0.070 & 0.000 & 0.000 & 0.000 &  \\
\rowcolor{Gray}
$1.37 \times 10^{0}$ & 3 & 14.6 & 1.5 & 0.909 & 0.903 & 0.913 & 0.000 & 0.000 & 0.000 & 0.019 & 0.020 & 0.016 & 0.072 & 0.077 & 0.071 & 0.000 & 0.000 & 0.000 &  \\
$1.37 \times 10^{0}$ & 3 & 32.8 & 1.0 & 0.981 & 0.981 & 0.982 & 0.000 & 0.000 & 0.000 & 0.003 & 0.003 & 0.003 & 0.016 & 0.016 & 0.015 & 0.000 & 0.000 & 0.000 &  \\
\rowcolor{Gray}
$1.37 \times 10^{0}$ & 3 & 32.8 & 1.5 & 0.993 & 0.992 & 0.993 & 0.000 & 0.000 & 0.000 & 0.001 & 0.002 & 0.001 & 0.006 & 0.006 & 0.006 & 0.000 & 0.000 & 0.000 &  \\
$1.37 \times 10^{0}$ & 4 & 8.2 & 1.0 & 0.334 & 0.304 & 0.358 & 0.000 & 0.000 & 0.000 & 0.051 & 0.053 & 0.051 & 0.615 & 0.642 & 0.590 & 0.000 & 0.001 & 0.001 &  \\
\rowcolor{Gray}
$1.37 \times 10^{0}$ & 4 & 8.2 & 1.5 & 0.361 & 0.343 & 0.378 & 0.000 & 0.000 & 0.000 & 0.057 & 0.060 & 0.056 & 0.582 & 0.596 & 0.566 & 0.000 & 0.001 & 0.000 &  \\
$1.37 \times 10^{0}$ & 4 & 14.6 & 1.0 & 0.577 & 0.543 & 0.591 & 0.000 & 0.000 & 0.000 & 0.038 & 0.038 & 0.035 & 0.385 & 0.419 & 0.374 & 0.000 & 0.000 & 0.000 &  \\
\rowcolor{Gray}
$1.37 \times 10^{0}$ & 4 & 14.6 & 1.5 & 0.562 & 0.529 & 0.569 & 0.000 & 0.000 & 0.000 & 0.059 & 0.060 & 0.055 & 0.379 & 0.406 & 0.375 & 0.000 & 0.005 & 0.001 &  \\
$1.37 \times 10^{0}$ & 4 & 32.8 & 1.0 & 0.885 & 0.869 & 0.895 & 0.000 & 0.000 & 0.000 & 0.014 & 0.014 & 0.013 & 0.101 & 0.115 & 0.092 & 0.000 & 0.002 & 0.000 &  \\
\rowcolor{Gray}
$1.37 \times 10^{0}$ & 4 & 32.8 & 1.5 & 0.889 & 0.872 & 0.897 & 0.000 & 0.000 & 0.000 & 0.011 & 0.013 & 0.011 & 0.100 & 0.112 & 0.092 & 0.000 & 0.003 & 0.000 &  \\
$1.37 \times 10^{0}$ & 5 & 8.2 & 1.0 & 0.094 & 0.073 & 0.113 & 0.000 & 0.000 & 0.000 & 0.055 & 0.057 & 0.056 & 0.851 & 0.866 & 0.831 & 0.000 & 0.004 & 0.000 &  \\
\rowcolor{Gray}
$1.37 \times 10^{0}$ & 5 & 8.2 & 1.5 & 0.130 & 0.098 & 0.149 & 0.000 & 0.000 & 0.000 & 0.068 & 0.069 & 0.067 & 0.802 & 0.827 & 0.782 & 0.000 & 0.006 & 0.002 &  \\
$1.37 \times 10^{0}$ & 5 & 14.6 & 1.0 & 0.249 & 0.197 & 0.272 & 0.000 & 0.000 & 0.000 & 0.050 & 0.053 & 0.049 & 0.700 & 0.742 & 0.678 & 0.001 & 0.008 & 0.001 &  \\
\rowcolor{Gray}
$1.37 \times 10^{0}$ & 5 & 14.6 & 1.5 & 0.235 & 0.200 & 0.271 & 0.000 & 0.000 & 0.000 & 0.063 & 0.065 & 0.060 & 0.701 & 0.729 & 0.668 & 0.001 & 0.006 & 0.001 &  \\
$1.37 \times 10^{0}$ & 5 & 32.8 & 1.0 & 0.715 & 0.662 & 0.726 & 0.000 & 0.000 & 0.000 & 0.016 & 0.017 & 0.016 & 0.269 & 0.304 & 0.256 & 0.000 & 0.017 & 0.002 &  \\
\rowcolor{Gray}
$1.37 \times 10^{0}$ & 5 & 32.8 & 1.5 & 0.700 & 0.646 & 0.719 & 0.000 & 0.000 & 0.000 & 0.019 & 0.024 & 0.018 & 0.280 & 0.310 & 0.261 & 0.001 & 0.020 & 0.002 &  \\
\bottomrule
\end{tabular}
\caption{ Outcomes of the $N_\mathrm{MC}=1000$ Monte Carlo realizations for various combinations of the grid parameters $t_\mathrm{V,1}$ (in units of yr, rounded to two decimal places), $N_\mathrm{p}$, $\beta$ (rounded to one decimal place) and $R_1$ (in units of $R_\mathrm{J}$). }
\label{table:MC_grid_results}
\end{table*}

\begin{table*}
\scriptsize
\begin{tabular}{cccccccccccccccccccc}
\toprule
& & & & \multicolumn{3}{c}{$f_\mathrm{no\, migration}$} & \multicolumn{3}{c}{$f_\mathrm{HJ}$} & \multicolumn{3}{c}{$f_\mathrm{TD}$} & \multicolumn{3}{c}{$f_{K_{ij}\leq0}$} & \multicolumn{3}{c}{$f_\mathrm{run\, time\,exceeded}$} \\
\cmidrule(l{2pt}r{2pt}){5-7} \cmidrule(l{2pt}r{2pt}){8-10} \cmidrule(l{2pt}r{2pt}){11-13}  \cmidrule(l{2pt}r{2pt}){14-16} \cmidrule(l{2pt}r{2pt}){17-19}
& & & & \multicolumn{2}{c}{$t_\mathrm{end} / \mathrm{Gyr}$} & & \multicolumn{2}{c}{$t_\mathrm{end} / \mathrm{Gyr}$} & & \multicolumn{2}{c}{$t_\mathrm{end} /\mathrm{Gyr}$} & & \multicolumn{2}{c}{$t_\mathrm{end} / \mathrm{Gyr}$} & & \multicolumn{2}{c}{$t_\mathrm{end} / \mathrm{Gyr}$} & \\
\cmidrule(l{2pt}r{2pt}){5-6} \cmidrule(l{2pt}r{2pt}){8-9} \cmidrule(l{2pt}r{2pt}){11-12}  \cmidrule(l{2pt}r{2pt}){14-15} \cmidrule(l{2pt}r{2pt}){17-18}
$t_\mathrm{V,1}/\mathrm{yr}$ & $N_\mathrm{p}$ & $\beta$ & $R_1/R_\mathrm{J}$ & 5 & 10 & $t_x$ & 5 & 10 & $t_x$ & 5 & 10 & $t_x$ & 5 & 10 & $t_x$ & 5 & 10 & $t_x$ \\
\midrule
$1.37 \times 10^{-2}$ & 3 & 8.2 & 1.0 & 0.835 & 0.818 & 0.846 & 0.002 & 0.004 & 0.002 & 0.028 & 0.032 & 0.027 & 0.135 & 0.146 & 0.125 & 0.000 & 0.000 & 0.000 &  \\
\rowcolor{Gray}
$1.37 \times 10^{-2}$ & 3 & 8.2 & 1.0 & 0.917 & 0.916 & 0.922 & 0.002 & 0.002 & 0.002 & 0.080 & 0.080 & 0.075 & $-$ & $-$ & $-$ & 0.001 & 0.002 & 0.001 &  \\$1.37 \times 10^{-2}$ & 3 & 14.6 & 1.0 & 0.913 & 0.908 & 0.916 & 0.001 & 0.001 & 0.000 & 0.021 & 0.022 & 0.021 & 0.065 & 0.069 & 0.063 & 0.000 & 0.000 & 0.000 &  \\
\rowcolor{Gray}
$1.37 \times 10^{-2}$ & 3 & 14.6 & 1.0 & 0.973 & 0.973 & 0.974 & 0.002 & 0.002 & 0.002 & 0.025 & 0.025 & 0.024 & $-$ & $-$ & $-$ & 0.000 & 0.000 & 0.000 &  \\
$1.37 \times 10^{-2}$ & 3 & 32.8 & 1.0 & 0.990 & 0.990 & 0.991 & 0.001 & 0.001 & 0.001 & 0.003 & 0.003 & 0.003 & 0.006 & 0.006 & 0.005 & 0.000 & 0.000 & 0.000 &  \\
\rowcolor{Gray}
$1.37 \times 10^{-2}$ & 3 & 32.8 & 1.0 & 0.996 & 0.996 & 0.996 & 0.000 & 0.000 & 0.000 & 0.004 & 0.004 & 0.004 & $-$ & $-$ & $-$ & 0.000 & 0.000 & 0.000 &  \\
$1.37 \times 10^{-2}$ & 4 & 8.2 & 1.0 & 0.305 & 0.289 & 0.335 & 0.000 & 0.000 & 0.000 & 0.068 & 0.071 & 0.068 & 0.627 & 0.638 & 0.596 & 0.000 & 0.002 & 0.001 &  \\
\rowcolor{Gray}
$1.37 \times 10^{-2}$ & 4 & 8.2 & 1.0 & 0.805 & 0.788 & 0.812 & 0.011 & 0.014 & 0.010 & 0.169 & 0.180 & 0.164 & $-$ & $-$ & $-$ & 0.015 & 0.018 & 0.014 &  \\
$1.37 \times 10^{-2}$ & 4 & 14.6 & 1.0 & 0.565 & 0.544 & 0.582 & 0.000 & 0.000 & 0.000 & 0.052 & 0.052 & 0.050 & 0.383 & 0.404 & 0.368 & 0.000 & 0.000 & 0.000 &  \\
\rowcolor{Gray}
$1.37 \times 10^{-2}$ & 4 & 14.6 & 1.0 & 0.879 & 0.868 & 0.880 & 0.005 & 0.005 & 0.004 & 0.106 & 0.111 & 0.106 & $-$ & $-$ & $-$ & 0.009 & 0.016 & 0.010 &  \\
$1.37 \times 10^{-2}$ & 4 & 32.8 & 1.0 & 0.887 & 0.872 & 0.897 & 0.000 & 0.000 & 0.000 & 0.008 & 0.008 & 0.007 & 0.105 & 0.118 & 0.096 & 0.000 & 0.002 & 0.000 &  \\
\rowcolor{Gray}
$1.37 \times 10^{-2}$ & 4 & 32.8 & 1.0 & 0.980 & 0.976 & 0.979 & 0.000 & 0.001 & 0.001 & 0.019 & 0.021 & 0.019 & $-$ & $-$ & $-$ & 0.001 & 0.002 & 0.001 &  \\
$1.37 \times 10^{-2}$ & 5 & 8.2 & 1.0 & 0.086 & 0.068 & 0.112 & 0.000 & 0.000 & 0.000 & 0.050 & 0.051 & 0.050 & 0.864 & 0.878 & 0.838 & 0.000 & 0.003 & 0.000 &  \\
\rowcolor{Gray}
$1.37 \times 10^{-2}$ & 5 & 8.2 & 1.0 & 0.637 & 0.586 & 0.641 & 0.008 & 0.009 & 0.007 & 0.306 & 0.319 & 0.298 & $-$ & $-$ & $-$ & 0.049 & 0.086 & 0.054 &  \\
$1.37 \times 10^{-2}$ & 5 & 14.6 & 1.0 & 0.243 & 0.201 & 0.266 & 0.001 & 0.001 & 0.001 & 0.059 & 0.059 & 0.057 & 0.696 & 0.729 & 0.675 & 0.001 & 0.010 & 0.001 &  \\
\rowcolor{Gray}
$1.37 \times 10^{-2}$ & 5 & 14.6 & 1.0 & 0.785 & 0.744 & 0.790 & 0.002 & 0.002 & 0.002 & 0.191 & 0.204 & 0.181 & $-$ & $-$ & $-$ & 0.022 & 0.050 & 0.027 &  \\
$1.37 \times 10^{-2}$ & 5 & 32.8 & 1.0 & 0.684 & 0.633 & 0.689 & 0.000 & 0.000 & 0.000 & 0.025 & 0.027 & 0.023 & 0.290 & 0.322 & 0.283 & 0.001 & 0.018 & 0.005 &  \\
\rowcolor{Gray}
$1.37 \times 10^{-2}$ & 5 & 32.8 & 1.0 & 0.954 & 0.913 & 0.952 & 0.001 & 0.002 & 0.001 & 0.044 & 0.048 & 0.039 & $-$ & $-$ & $-$ & 0.001 & 0.037 & 0.008 &  \\
\bottomrule
\end{tabular}
\caption{ Similar to Table\,\ref{table:MC_grid_results}, here showing results from a subset of simulations without the $K_{ij}\leq 0$ stopping condition which can be recognized from the entries with $f_{K_{ij}\leq 0}$ marked as $-$. For convenience, the corresponding entries with the $K_{ij}\leq 0$ stopping condition enabled are included as well (repeated from Table\,\ref{table:MC_grid_results}). }
\label{table:MC_grid_results_K_ij}
\end{table*}

\subsection{Results}
\label{sect:pop_syn:results}
\subsubsection{Overview}
\label{sect:pop_syn:results:overview}
Our results are summarized in Table \ref{table:MC_grid_results}. For each combination of $t_\mathrm{V,1}$, $N_\mathrm{p}$, $\beta$ and $R_1$, we list the fractions with respect to the $N_\mathrm{MC}=1000$ Monte Carlo-sampled systems of the following outcomes, closely related to the stopping conditions.
\begin{enumerate}
\item No migration occurred, i.e. the final orbital period $P_1>100\, \mathrm{d}$ ($f_\mathrm{no\,migration}$).
\item A HJ was formed ($f_\mathrm{HJ}$; WJs: see below).
\item The innermost planet was tidally disrupted ($f_\mathrm{TD}$).
\item $K_{ij}\leq 0$ occurred for any orbit pair ($f_{K_{ij}\leq 0}$).
\item The maximum run time of 12 CPU hours was exceeded ($f_\mathrm{run\, time\,exceeded}$).
\end{enumerate}
These fractions are given after either 5 or 10 Gyr of integration, or by sampling a random time between 100 Myr and 10 Gyr for each of the systems, corresponding to a constant star formation rate (indicated with $t_x$ in the table; output times were separated by 100 Myr). For $t_\mathrm{V,1} \approx 1.4 \times 10^{-2} \, \mathrm{yr}$ and $R_1 = 1 \, R_\mathrm{J}$, we also carried out simulations without the $K_{ij}\leq 0$ stopping condition. The results from the latter simulations are given in Table\,\ref{table:MC_grid_results_K_ij}, and can be recognized in that table from the entries with $f_{K_{ij}\leq 0}$ marked as $-$. 

The fractions of HJs formed are typically low; the largest fraction is 0.023, obtained after 10 Gyr for the set of simulations with $t_\mathrm{V,1} \approx 1.4 \times 10^{-3} \, \mathrm{yr}$, $N_\mathrm{p}=3$, $\beta\approx 8.2$ and $R_1 = 1.5\,R_\mathrm{J}$. In contrast, the fraction of tidal disruptions is larger, typically a few per cent, and reaching values of $\approx 0.2-0.3$ for $N_\mathrm{p}=5$ if the $K_{ij}\leq 0$ stopping condition is not imposed (cf. Table\,\ref{table:MC_grid_results_K_ij}). 

The number of WJs (defined as planets with an orbital period between 10 and 100 d at a given time) is even smaller than the number of HJs, and the associated fractions are not included in Tables\,\ref{table:MC_grid_results} and \ref{table:MC_grid_results_K_ij}. Among the 72,000 integrations in Table\,\ref{table:MC_grid_results}, 186 HJs were formed after 10 Gyr, whereas the number of WJs at that time is 11 (the number of WJs and HJs at 5 Gyr is seven and 166, respectively). Moreover, the semimajor axes of the WJs are $\approx 0.1\,\mathrm{AU}$ (cf. \F\,\ref{fig:slr_distributions_dependence_on_all_pars_MC02d}), i.e. the WJs are on the `hot' end of the WJ spectrum, and near the (not well-defined) boundary between WJs and HJs. 

For values of $t_\mathrm{V,1} \gtrsim 1.4 \times 10^{-2}\, \mathrm{yr}$, i.e. for relatively weak tidal dissipation strength in the innermost planet, no HJs are formed at all, for any of the combinations of the grid parameters. Note that the number of Monte Carlo realizations per parameter combination was limited to $N_\mathrm{MC}=1000$, implying that the HJ fractions could be less than 0.001, but nonzero. Also, uncertainties associated with the stopping condition $K_{ij}\leq 0$ and premature terminations of the integrations because of the exceeding of the maximum run time, should be taken into account. These are discussed in more detail in \S\,\ref{sect:discussion:uncertainties}. 

Note that the cumulative `non-migrating' fractions in Table \ref{table:MC_grid_results} are typically lower compared to the fraction of HJs found in \S\,\ref{sect:ver}. This can be attributed to the high initial eccentricities and inclinations that were assumed in \S\,\ref{sect:ver}, of $\approx 0.5$, whereas they were typically lower in the population synthesis.

\subsubsection{Final orbital period distributions}
\label{sect:pop_syn:results:period}

\begin{figure}
\center
\includegraphics[scale = 0.45, trim = 0mm 0mm 0mm 0mm]{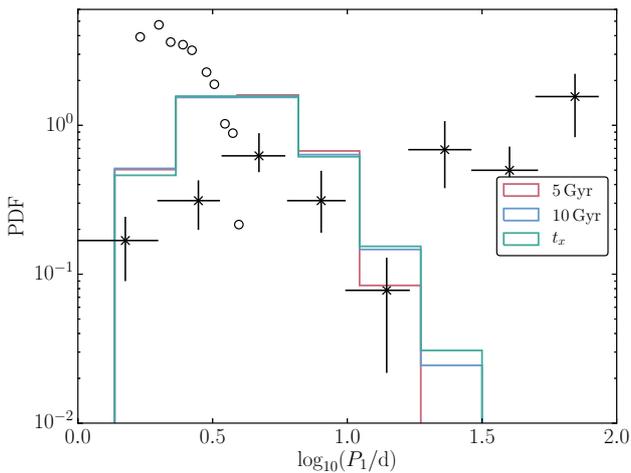}
\caption{\small Solid lines: the distributions of the orbital period of the innermost planet at 5 and 10 Gyr, and at a random time ($t_x$). Crosses: observations from \citet{2016A&A...587A..64S}; error bars are indicated with black lines. Open circles: distribution from fig. 23 of \citet{2016MNRAS.456.3671A} for $M_\mathrm{p}=1\,M_\mathrm{J}$ and $\chi=100$. Distributions are normalized to unit total area. }
\label{fig:sma_distributions_combined_MC02d}
\end{figure}

In \F\,\ref{fig:sma_distributions_combined_MC02d}, we show the distributions of the orbital periods of the innermost planet at various times, combining results from all parameter combinations. With the solid lines, we show the distributions for the `non-disruptive' systems in which a WJ or HJ was formed, or no migration occurred (i.e. excluding outcomes iii through v from \S\,\ref{sect:pop_syn:results:overview}). We consider the distributions after 5 Gyr (red line), 10 Gyr (blue line), and assuming a random time for each system between 100 Myr and 10 Gyr ($t_x$; green line). The crosses show the (unbiased) observed distribution from \citet{2016A&A...587A..64S}; error bars are indicated with black lines. Open circles show the distribution from fig. 23 of \citet{2016MNRAS.456.3671A} for $m_1 \equiv M_\mathrm{p}=1\,M_\mathrm{J}$ and $\chi=100$, where $\chi \equiv 10 \, \tau_1/\mathrm{s}$ and $\tau_1$ is the tidal time lag of the innermost planet (cf. table 1 of the latter paper). With $k_\mathrm{AM,1} = 0.25$ (cf. Table\,\ref{table:IC}), $m_1 = 1\,M_\mathrm{J}$ and $R_1 = 1 \, R_\mathrm{J}$, $\chi=100$ or $\tau_1 = 10 \, \mathrm{s}$ corresponds to a viscous time-scale of $\approx 0.082 \, \mathrm{yr}$ (or a viscous time-scale of $\approx 0.28 \, \mathrm{yr}$ for $R_1 = 1.5 \, R_\mathrm{J}$). 

The simulated orbital period distribution is peaked around $\mathrm{log}_{10}(P_1/\mathrm{d}) \sim 0.7$, or $P_1 \sim 5 \, \mathrm{d}$. The location of the peak in the simulations is consistent with observations, which show a peak at the same orbital period. Compared to \citet{2016MNRAS.456.3671A}, who considered high-$e$ migration in stellar binaries, our orbital period distribution is wider and peaked at longer periods ($\sim 5$ d versus $\sim 2$ d). This is likely not only due to the (fundamental) difference in the orbital configuration (an inclined three-body system versus a mildly inclined multiplanet system with three to five planets), but also other parameters such as the viscous time-scale and in particular the planetary radius (cf. \S\,\ref{sect:pop_syn:results:grid_dependence}).

Similarly to previous studies of high-$e$ migration in other contexts (e.g. \citealt{2016ApJ...829..132P,2016AJ....152..174A}), the simulations fail to produce the large observed population of WJs in the region between $\sim 10$ and 100 d. 

\begin{figure}
\center
\includegraphics[scale = 0.45, trim = 10mm 0mm 0mm 0mm]{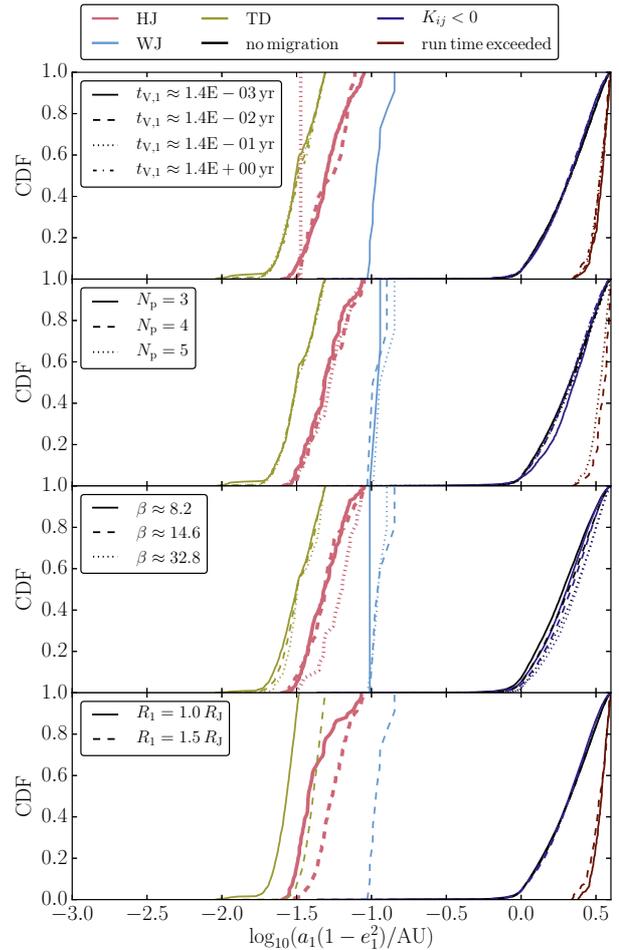}
\caption{\small The innermost orbit semilatus rectum distributions after 10 Gyr (or after a stopping condition was met) plotted for different slices of the parameter space. In each panel, distributions are shown for all parameters combined, except one (different line styles). The varying parameters are $t_\mathrm{V,1}$, $N_\mathrm{p}$, $\beta$ and $R_1$ in the top through bottom panels, respectively. Different types of systems are indicated with different colours: systems with no migration (black), HJs (light red), WJs (light blue), tidally disrupted inner planets (yellow), $K_{ij}\leq 0$ (dark blue) or exceeding of run time (dark red). }
\label{fig:slr_distributions_dependence_on_all_pars_MC02d}
\end{figure}

\subsubsection{Dependence on the grid parameters}
\label{sect:pop_syn:results:grid_dependence}
In \F\,\ref{fig:sma_distributions_combined_MC02d}, we combined results from all parameters. Here, we consider in more detail the dependence of the results on the parameters individually. 

In \F\,\ref{fig:slr_distributions_dependence_on_all_pars_MC02d}, the innermost orbit semilatus rectum distributions after 10 Gyr (or after a stopping condition was met) are plotted for different slices of the parameter space. Considering tidal evolution only, the final semimajor axis (at the moment of circularisation) is expected to be equal to the semilatus rectum. In each panel, distributions are shown for all parameters combined, except one (different line styles). In addition, we distinguish between the different types of systems: no migration (black), HJs (light red), WJs (light blue), tidally disrupted inner planets (yellow), $K_{ij}\leq 0$ (dark blue) or exceeding of run time (dark red).

In the top panel, we show the dependence on $t_\mathrm{V,1}$. The distributions for the non-HJ and non-WJ forming systems are essentially independent of $t_\mathrm{V,1}$. HJs and WJs are only formed for viscous time-scales of $\lesssim 1.4\times 10^{-2}\,\mathrm{yr}$ and $\approx 1.4\times 10^{-3}\,\mathrm{yr}$, respectively. The requirement of highly efficient tides for HJ production was also found for other high-$e$ migration scenarios, in particular in stellar binaries \citep{2015ApJ...799...27P}. 

The dependence on the number of planets and $\beta$ is shown in the second and third panels of \F\,\ref{fig:slr_distributions_dependence_on_all_pars_MC02d}, respectively. Despite the expected propensity of exciting higher eccentricities with a larger number of planets and/or smaller $\beta$, the dependence of the semilatus rectum distributions on these parameters is not markably strong.

The dependence on the radius of the innermost planet is shown in the bottom panel of \F\,\ref{fig:slr_distributions_dependence_on_all_pars_MC02d}. For HJs, the final semilatus rectum is smaller for smaller radii (see e.g. equation 3 of \citealt{2011ApJ...735..109W}). For the tidally disrupted planets, a larger radius corresponds to a larger tidal disruption radius (cf. equation~\ref{eq:r_t}), and therefore a larger semilatus rectum at the moment of disruption.

\subsubsection{Dependence on the initial orbital properties}
\label{sect:pop_syn:results:orbit_dep}

\begin{figure}
\center
\includegraphics[scale = 0.45, trim = 0mm 0mm 0mm 0mm]{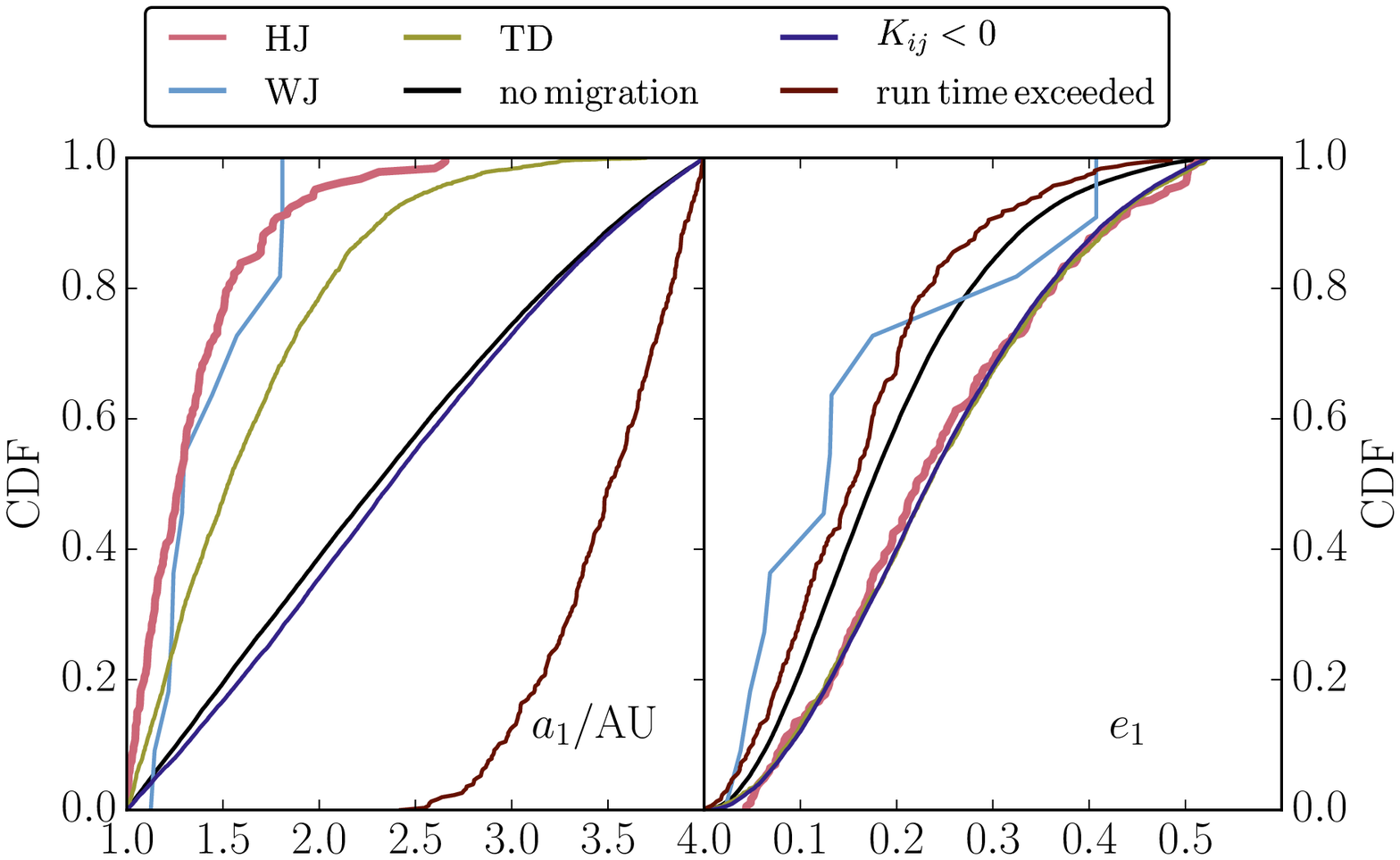}
\includegraphics[scale = 0.45, trim = 0mm 0mm 0mm 0mm]{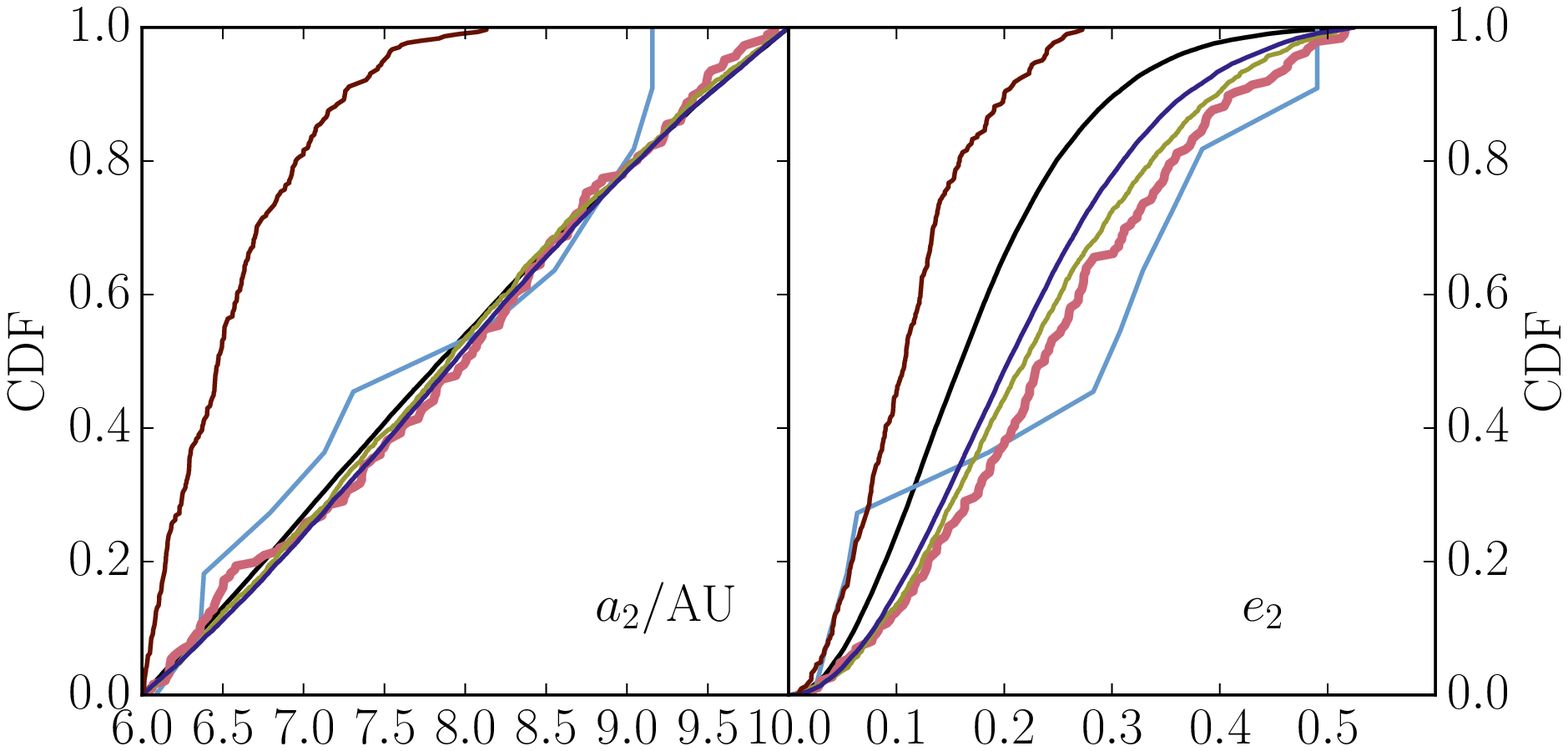}
\includegraphics[scale = 0.45, trim = 0mm 0mm 0mm 0mm]{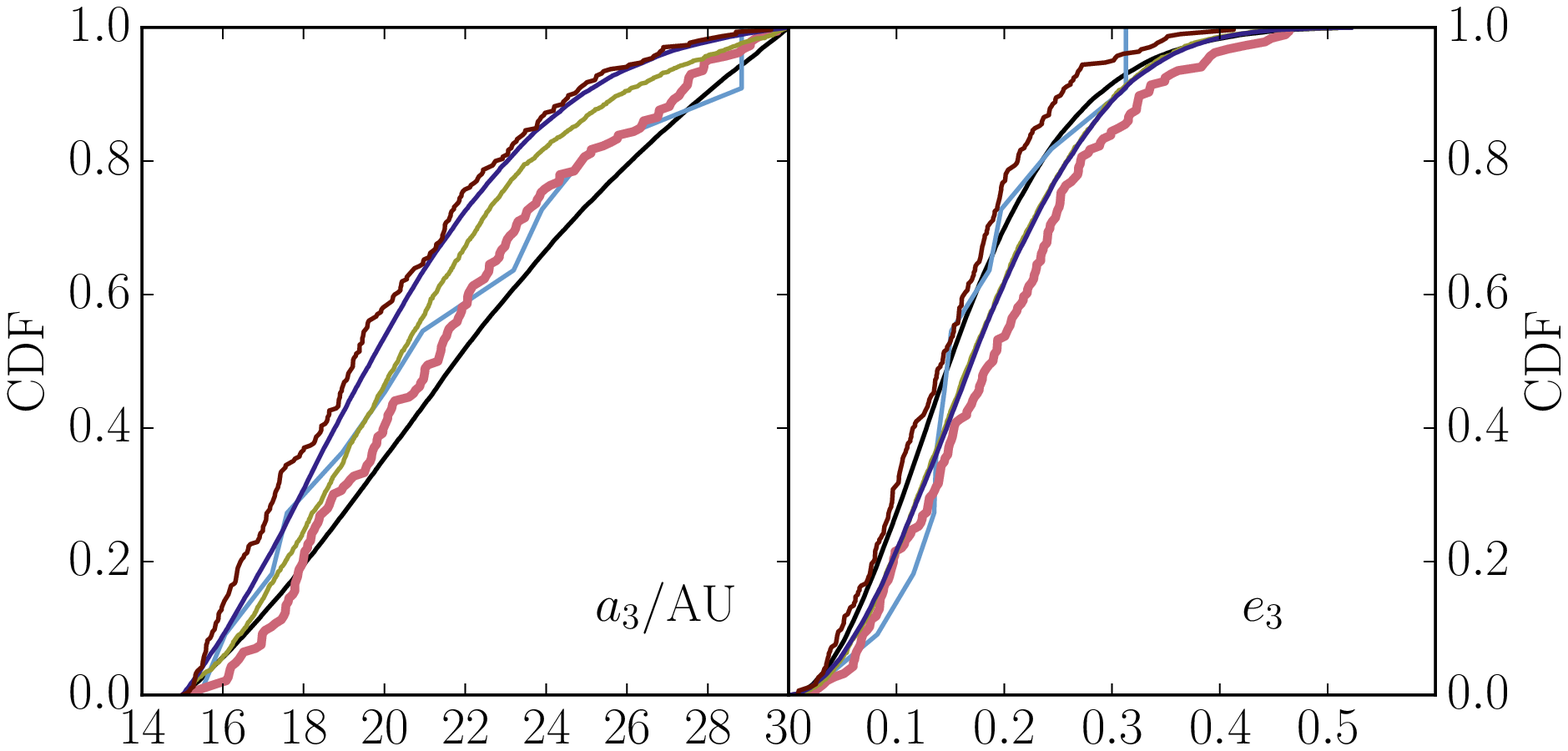}
\caption{\small The initial distributions of the semimajor axes (left column) and the eccentricities (right column) for the various outcomes of the Monte Carlo simulations. Colours indicate systems with no migration (black), HJs (light red), WJs (light blue), tidally disrupted inner planets (yellow), $K_{ij}\leq 0$ (dark blue) or exceeding of run time (dark red). }
\label{fig:MC_initial_smas_es}
\end{figure}
In \F\,\ref{fig:MC_initial_smas_es}, we show how the various outcomes in our simulations depend on the initial semimajor axes (left column) and the eccentricities (right column), for the three innermost orbits. We recall that the semimajor axes were sampled linearly from fixed ranges, whereas the eccentricities were sampled from a Rayleigh distribution with various widths expressed by $\beta$ (cf. \S\,\ref{sect:pop_syn:IC}). 

The largest differences in the initial semimajor axes between the various outcomes are apparent in the innermost orbit. The distribution of the initial $a_1$ for the HJ and tidal disruption systems is skewed towards small values compared to the other systems, with $a_1\lesssim 2\,\mathrm{AU}$ for most ($\approx 0.95$ and $\approx 0.8$, respectively) of the systems. This can be attributed to two effects. For the typical $a_2$ and $a_3$ in the simulations, a small $a_1$ implies a larger commensurability between the LK time-scales associated with orbit pairs (1,2) and (2,3), and therefore more likely chaotic evolution and higher eccentricities (e.g. \citealt{2015MNRAS.449.4221H}). Also, the required eccentricities for small pericentre distances (important for tidal dissipation or tidal disruption) are lower for smaller semimajor axes. 

The systems with exceeded run times (dark red lines) preferentially have large $a_1$, whereas $a_2$ is preferentially small. These systems are unlikely to result in HJs (cf. \S\,\ref{sect:discussion:uncertainties}). With regard to the other orbits and outcomes, no strong differences can be discerned in the initial distributions of the semimajor axes. 

HJ and tidal disruption systems typically show a preference for initially higher values of the eccentricities, notably $e_2$. Otherwise, there is no strong dependence on the initial eccentricities.

\begin{figure}
\center
\includegraphics[scale = 0.45, trim = 0mm 0mm 0mm 0mm]{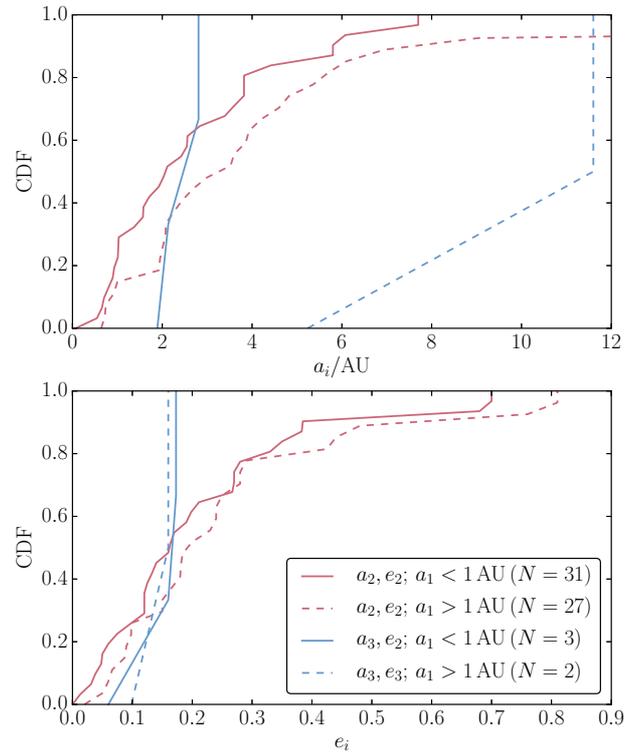}
\caption{\small Observed distributions of the semimajor axes (top panel) and the eccentricities (bottom panel) for planets with $m_1 \sin(i)>0.1\,M_\mathrm{J}$ obtained from the {\it Open Exoplanet Catalogue}. We consider multiplanet systems with at least two observed planets, and make a distinction between systems with the innermost observed orbit $a_1<1\,\mathrm{AU}$ (solid lines) and $a_1>1\,\mathrm{AU}$ (dashed lines). }
\label{fig:observed_companion_distributions}
\end{figure}

Observations of companion planets to HJs are currently still strongly limited. In the latest surveys, detections of Jupiter-mass planets are only 100\% complete for planets out to $\approx 10 \, \mathrm{AU}$ \citep{2016ApJ...821...89B}. In our simulations, except for the innermost two planets, the orbits span a much larger range in semimajor axis. Despite the incompleteness of the observations, we show in \F\,\ref{fig:observed_companion_distributions} observed distributions of the semimajor axes (top panel) and the eccentricities (bottom panel) for planets with $m_1 \sin(i)>0.1\,M_\mathrm{J}$ obtained from the {\it Open Exoplanet Catalogue} (\citealt{2012arXiv1211.7121R}; \href{https://github.com/OpenExoplanetCatalogue}{https://github.com/OpenExoplanetCatalogue}). We consider multiplanet systems with at least two observed planets, and make a distinction between systems with the innermost observed orbit $a_1<1\,\mathrm{AU}$ (solid lines) and $a_1>1\,\mathrm{AU}$ (dashed lines). Some difference can be seen in the distributions of $a_2$ for the two populations with $a_1<1\,\mathrm{AU}$ and $a_1>1\,\mathrm{AU}$: $a_2$ tends to be smaller for the former population. This trend is not reflected in our simulations (cf. the second row of \F\,\ref{fig:MC_initial_smas_es}). With regard to $a_3$, there seems to be a large difference in the distribution of $a_3$ for the two populations. However, our observational sample only includes five systems with at least three planets, so with this low number of systems this difference cannot be considered significant. 

Another, more theoretically oriented, quantity is the angular-momentum deficit (AMD). The AMD is defined as
\begin{align}
\label{eq:AMD_def}
\mathrm{AMD} = \sum_{i=1}^{N_\mathrm{p}} m_i\sqrt{a_i} \left [ 1 - \sqrt{1-e_i^2} \cos(i_{\mathrm{var},i}) \right ],
\end{align}
where
$i_{\mathrm{var},i}$ is the inclination with respect to the invariable plane, i.e. the plane perpendicular to the total orbital angular-momentum vector of the system. In terms of the AMD, high eccentricities and/or chaotic motion can be achieved if $\mathrm{AMD} \gtrsim m_1 \sqrt{a_1}$ \citep{2011ApJ...735..109W,2011ApJ...739...31L,2014PNAS..11112610L}. 

\begin{figure}
\center
\includegraphics[scale = 0.45, trim = 0mm 0mm 0mm 0mm]{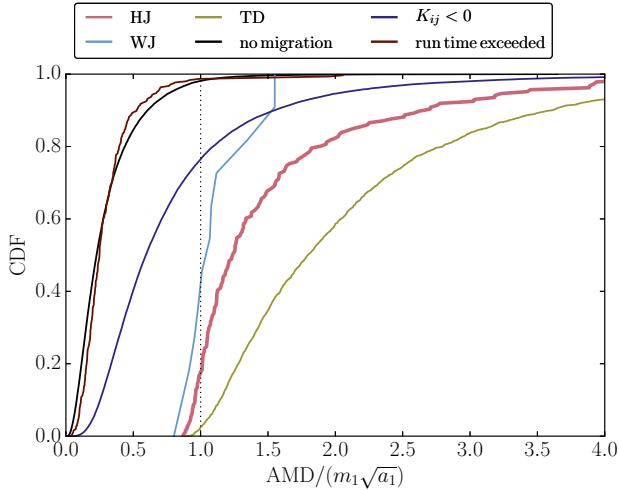}
\caption{\small The initial distributions of the AMD for the various outcomes of the Monte Carlo simulations. Colours indicate systems with no migration (black), HJs (light red), WJs (light blue), tidally disrupted inner planets (yellow), orbit crossings (dark blue) or exceeding of run time (dark red). The black vertical dotted line indicated $\mathrm{AMD} = m_1 \sqrt{a_1}$.}
\label{fig:MC_AMD_cases_MC02d}
\end{figure}

In our simulations, there is indeed a strong dependence on the AMD. In \F\,\ref{fig:MC_AMD_cases_MC02d}, we show the initial distributions of the AMD for the various outcomes. Of the non-migrating systems (black line), nearly all ($\approx 0.99$) have an AMD which is $<m_1\sqrt{a_1}$. In contrast, virtually all tidally disrupted systems (yellow line) have an $\mathrm{AMD} \gtrsim m_1\sqrt{a_1}$, and the HJ systems (light red line) have an $\mathrm{AMD} \gtrsim 0.8 \, m_1\sqrt{a_1}$, with the majority ($\sim 0.8$) of systems having $\mathrm{AMD}> m_1\sqrt{a_1}$. 

The systems in which $K_{ij}\leq 0$ occurred (dark blue line) have higher AMD compared to the non-migrating systems, which can be attributed to the higher eccentricities attained with higher AMDs, therefore more likely leading to $K_{ij}\leq 0$. Nonetheless, $\approx 0.75$ of these systems have $<m_1\sqrt{a_1}$, indicating that the majority of systems with $K_{ij}\leq 0$ would not have produced HJs or tidally disrupted planets if the stopping condition at $K_{ij}\leq 0$ had not been imposed. This is consistent with the result that the HJ fractions are not strongly affected in the runs without this stopping condition (cf. Table\,\ref{table:MC_grid_results_K_ij} and \S\,\ref{sect:pop_syn:results:overview}).

There are distinct differences with respect to the AMD between the HJ and tidal disruption systems. Many ($\approx 0.7$) of the HJs have $\mathrm{AMD}<1.5\,m_1\sqrt{a_1}$, whereas for the tidal disruption systems this fraction is markedly lower, $\approx 0.4$. The preference for the tidal disruption systems for higher AMDs can be explained by the higher eccentricities reached in the innermost orbit, therefore more likely resulting in the (immediate) tidal disruption of the innermost planet, rather than tidal dissipation, which requires a certain amount of time to dissipate energy and reduce the eccentricity. This implies that there is a `window' for producing HJs through secular evolution: the AMD should be large enough to excite high eccentricities, but small enough to prevent violent excitation of the eccentricities leading to tidal disruption before tidal dissipation can be effective. 

\begin{figure}
\center
\includegraphics[scale = 0.45, trim = 0mm 0mm 0mm 0mm]{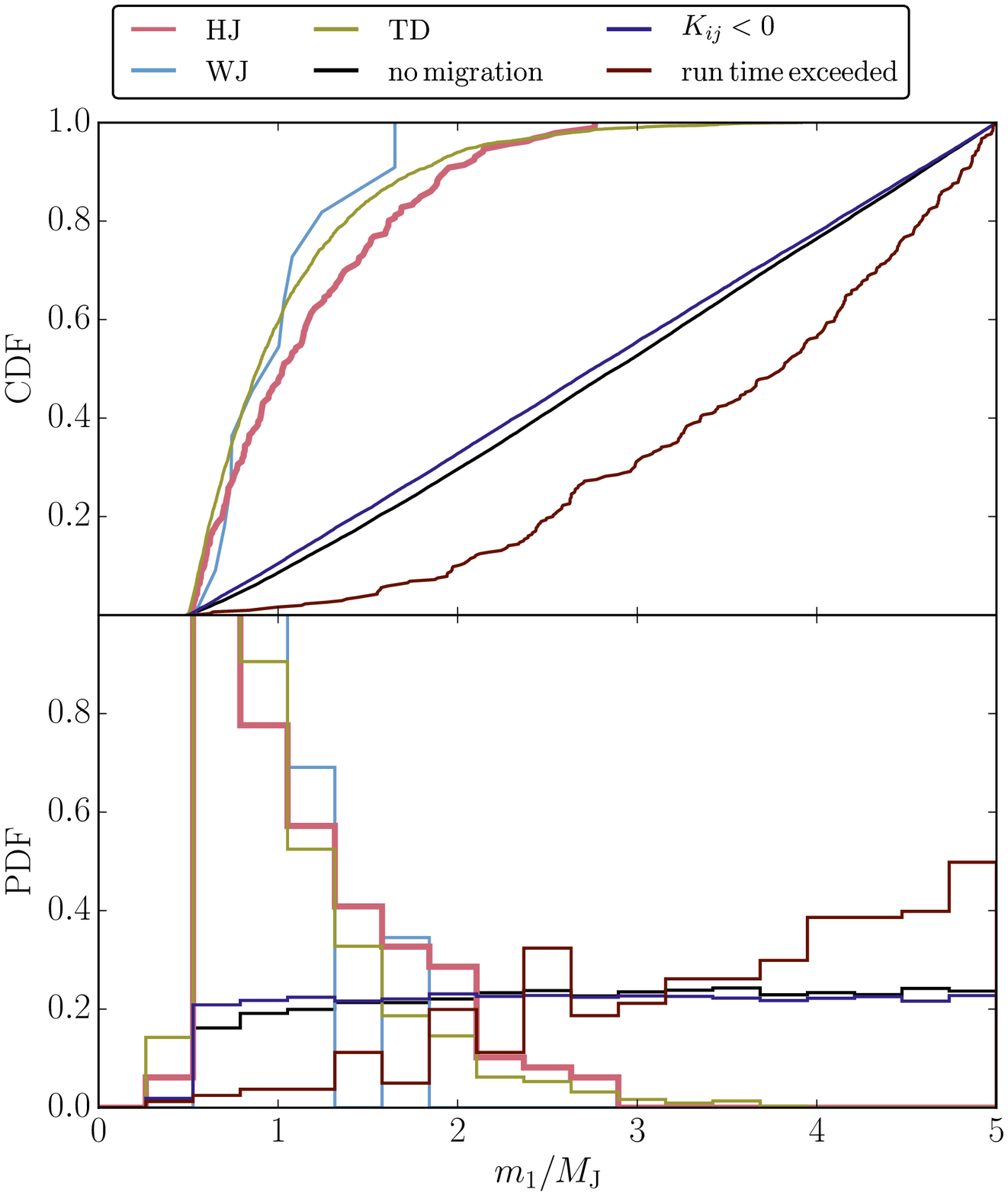}
\includegraphics[scale = 0.45, trim = 0mm 0mm 0mm 0mm]{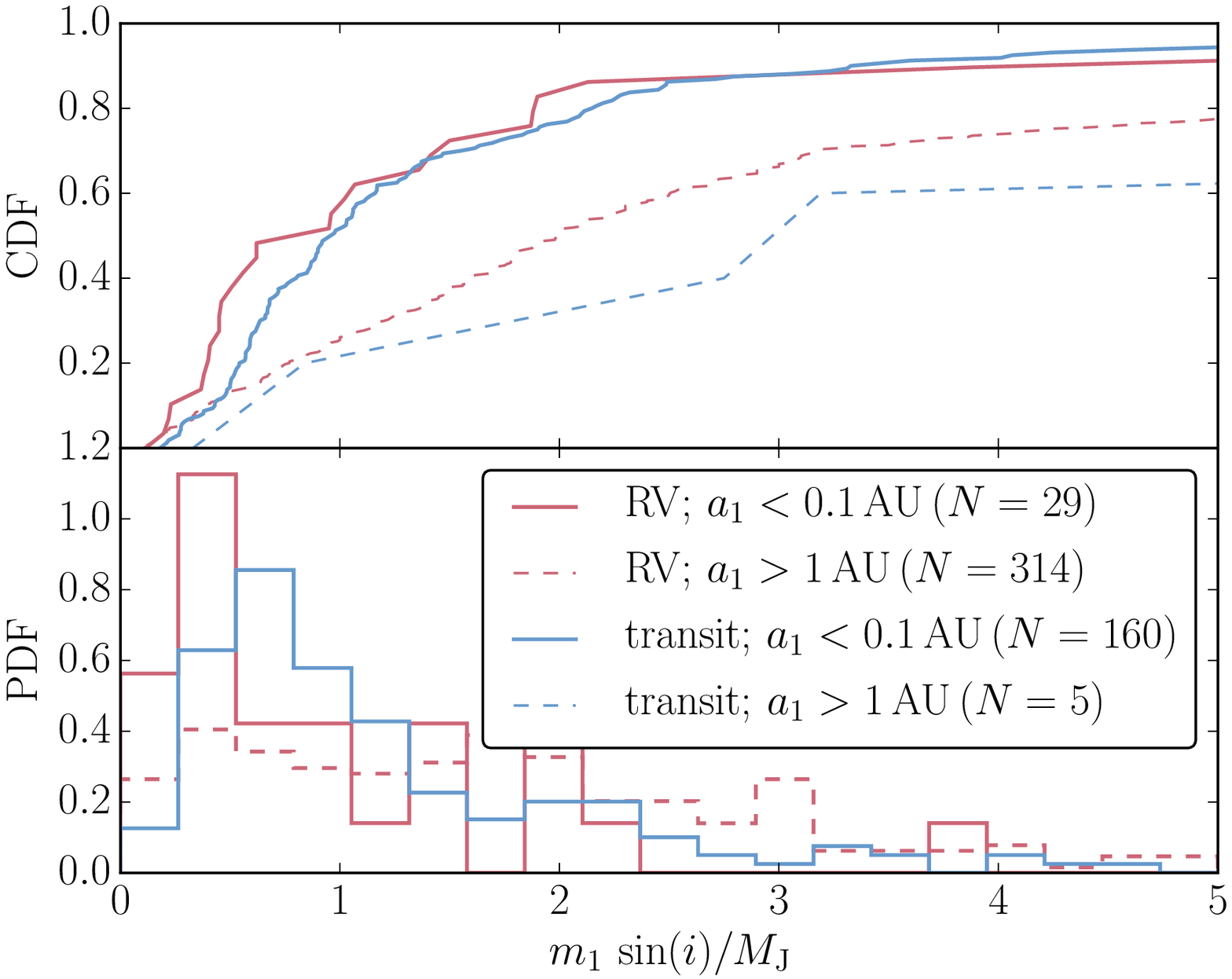}
\caption{\small Top panels: distributions of the initial mass of the innermost planet for the Monte Carlo simulations with all grid parameters combined. Different types of systems are indicated with different colours: systems with no migration (black), HJs (light red), WJs (blue), tidally disrupted inner planets (yellow), $K_{ij}\leq 0$ (dark blue) or exceeding of run time (dark red). Bottom panels: the observed mass distributions of planets with $m_1 \sin(i)>0.1\,M_\mathrm{J}$ (obtained from the {\it Open Exoplanet Catalogue}). Two discovery methods are included: RV (red) and transit (blue). Also, a distinction is made between the semimajor axis: $a_1<0.1\,\mathrm{AU}$ (solid lines) and $a_1>1\,\mathrm{AU}$ (dashed lines). }
\label{fig:m_1_distributions_combined_MC02d}
\end{figure}

\subsubsection{Mass dependence}
\label{sect:pop_syn:results:mass}
The orbits of lower-mass planets carry less orbital angular momentum compared to higher-mass counterparts, making the former more susceptible to angular-momentum exchanges with orbits of other planets. Therefore, secular eccentricity excitation is expected to be more pronounced if the outer planets are more massive than the innermost planet. In our simulations, we assumed a flat distribution of the planetary masses between 0.5 and 5 $M_\mathrm{J}$.

In the top panels of \F\,\ref{fig:m_1_distributions_combined_MC02d}, we show the distributions of the mass of the innermost planet for the various outcomes in the simulations. For the non-migrating systems, the mass distribution is consistent with a flat distribution, reflecting the initial distribution, and showing no mass preference. HJ and tidal disruption systems show different mass distributions. For the latter groups, there is a preference for lower-mass planets, with $m_1\lesssim 1\,M_\mathrm{J}$ for $\sim 0.4$ and $\sim 0.5$ of the systems, respectively. There are few HJs and tidal disruption systems with $m_1\gtrsim 2\,M_\mathrm{J}$. This implies a clear quantitative prediction for high-$e$ migration in multiplanet systems, and which was given qualitatively in \citet{2011ApJ...735..109W}: the HJ should have a typical (median) mass of $\sim 1 \,M_\mathrm{J}$, and not be more massive than $\sim 2 \,M_\mathrm{J}$.

Observations show a deficit of massive HJs \citep{2002ApJ...568L.113Z,2007ARA&A..45..397U}, which seems consistent with the above prediction. More quantitatively, the observed mass distributions of planets with $m_1 \sin(i)>0.1\,M_\mathrm{J}$ (obtained from the {\it Open Exoplanet Catalogue}) are shown in the bottom panels of \F\,\ref{fig:m_1_distributions_combined_MC02d}. We made a distinction between discovery method (RV or transit) and semimajor axis ($a_1<0.1\,\mathrm{AU}$ and $a_1>1\,\mathrm{AU}$). The planets from the RV observations within 0.1 AU are typically of lower mass compared to planets at $> 1 \, \mathrm{AU}$. This trend is consistent with the predictions as described above. However, one should be cautious when ascribing the observed mass difference to secular evolution alone, given that the latter unlikely produces all HJs, and the observed mass distribution is also likely affected by other processes, such as primordial `mass segregation', whereas in the simulations we assumed an initially flat distribution. Moreover, the RV observations are biased, because planets at $>1\,\mathrm{AU}$ are more easily detected if they are more massive.

\begin{figure}
\center
\includegraphics[scale = 0.45, trim = 10mm 0mm 0mm 0mm]{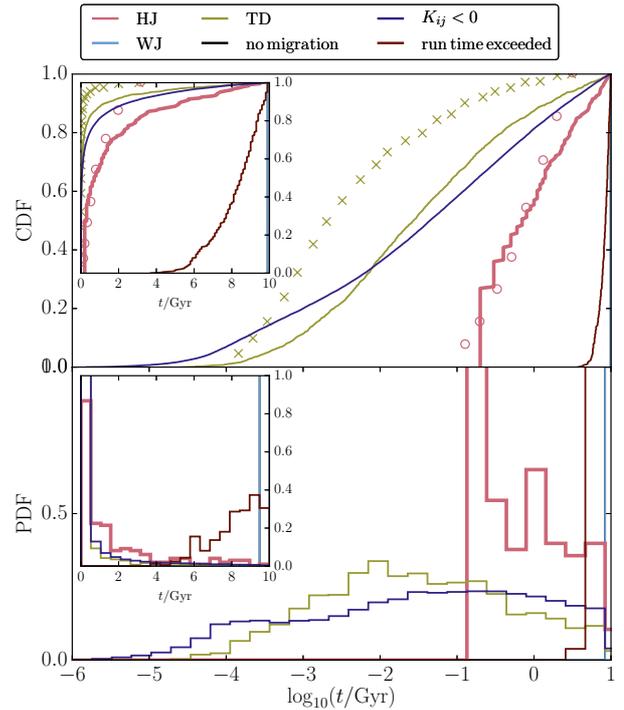}
\caption{\small The final times associated with the various outcomes in our population synthesis (with all parameters combined). For systems in which a stopping condition occurred, these final times are the age of the system when the integration was stopped. In the case of HJs (light red lines), these times are the times of HJ formation. Different types of systems are indicated with different colours: systems with no migration (black), HJs (light red), WJs (blue), tidally disrupted inner planets (yellow), $K_{ij}\leq 0$ (dark blue) or exceeding of run time (dark red). The same data, shown using linear scales on the abscissa, are shown in the insets. Red open circles and yellow crosses: data for HJs and tidal disruptions, respectively, from the second panel of fig. 22 ($M_\mathrm{p}=1\,M_\mathrm{J}$) of \citet{2016MNRAS.456.3671A}. }
\label{fig:end_times_MC02d}
\end{figure}

\subsubsection{HJ formation times}
\label{sect:pop_syn:results:times}
As mentioned in \S\,\ref{sect:introduction}, secular evolution in multiplanet systems typically occurs on long time-scales of the order of Gyr, implying that HJs formed through this mechanism could have been deposited at their current orbit at late stages in the MS lifetime of the host star. In \F\,\ref{fig:end_times_MC02d}, we show the `final' times associated with the various outcomes in our population synthesis (with all parameters combined). For systems in which a stopping condition occurred, these final times are the age of the system when the integration was stopped. In the case of HJs (light red lines), these times are the times of HJ formation, as defined in \S\,\ref{sect:pop_syn:SC}. 

The HJs in our simulations are indeed formed late, with a median formation time of $\approx 1 \, \mathrm{Gyr}$, and with a fraction $\approx 0.1$ of the HJs formed after $\approx 6 \, \mathrm{Gyr}$. These times are much longer compared to times associated with high-$e$ migration due to close encounters \citep{1996Sci...274..954R,2008ApJ...686..580C,2008ApJ...686..621F,2008ApJ...686..603J}, and somewhat longer compared to those typically found for high-$e$ migration in stellar binaries (\citealt{2016MNRAS.456.3671A}; cf. the red open circles in \F\,\ref{fig:end_times_MC02d}). In contrast, tidal disruptions occur much earlier, with $\approx 0.85$ of the disruptions occurring before 1 Gyr. We note that the peak around 100 Myr in the HJ formation times arises from the systems with the shortest viscous time-scale of $\approx 1.4 \times 10^{-3} \, \mathrm{yr}$.

\begin{figure}
\center
\includegraphics[scale = 0.45, trim = 10mm 0mm 0mm 0mm]{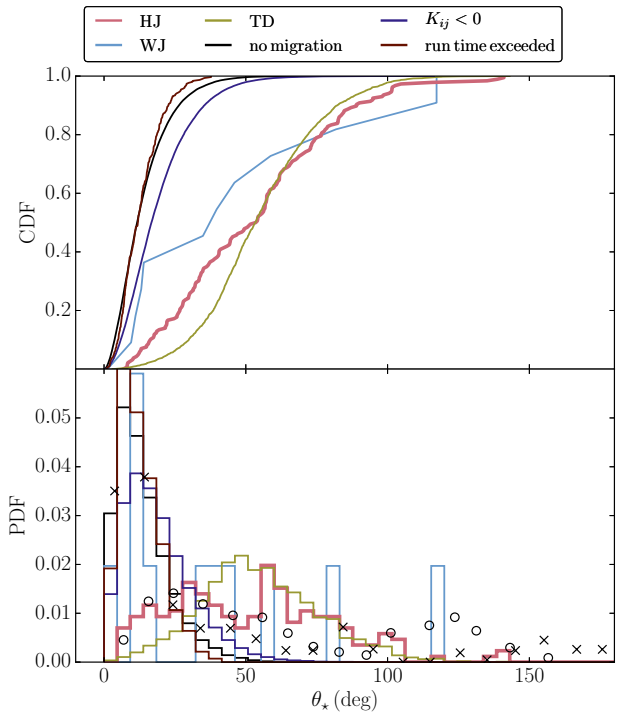}
\caption{\small Distributions after 10 Gyr (or after a stopping condition) of the stellar obliquity $\theta_\star$ for the Monte Carlo simulations with all grid parameters combined. As in \F\,\ref{fig:slr_distributions_dependence_on_all_pars_MC02d}, different types of systems are indicated with different colours: systems with no migration (black), HJs (light red), WJs (blue), tidally disrupted inner planets (yellow), $K_{ij}\leq 0$ (dark blue) or exceeding of run time (dark red). Open circles: distribution from fig. 24 of \citet{2016MNRAS.456.3671A} for $M_\mathrm{p}=1\,M_\mathrm{J}$ and $\chi=100$. Crosses: observed projected obliquity distribution, adopted from \citet{2014PNAS..11112610L}. Note that observed (i.e. projected) obliquities are shifted to lower angles relative to the intrinsic obliquities. The PDFs are normalized to unit total area. }
\label{fig:spin}
\end{figure}

\subsubsection{Stellar obliquity}
\label{sect:pop_syn:results:spin}
In \F\,\ref{fig:spin}, we show, for all grid parameters combined, the distributions after 10 Gyr (or after a stopping condition) of the stellar obliquity, i.e. the angle between the stellar spin and the orbit of the innermost planet. In the simulations, spin-orbit coupling was included taking into account the spin directions of both the star and the innermost planet (cf. \S\,\ref{sect:methods:tides}). Initially, the stellar spin and innermost orbit were assumed to be aligned, i.e. zero obliquity was assumed (cf. Table\,\ref{table:IC}).

For high-$e$ migration scenarios with three bodies (or, more generally, LK cycles with tidal friction), it has been well established that the obliquity for HJs should be clustered around $40^\circ$ and $130^\circ$ \citep{2007ApJ...669.1298F,2014ApJ...793..137N,2016MNRAS.456.3671A}. This can be explained intuitively by noting that during LK cycles, the eccentricity maxima occur at mutual inclinations of $\approx 40^\circ$ or $130^\circ$, and mutual inclination tends to be `locked' after the onset of strong tidal dissipation (this is assuming that the stellar spin vector itself does not change due to spin-planet orbit coupling; see \citealt{2014Sci...345.1317S,2015MNRAS.448.1821S,2016MNRAS.456.3671A}). Observations have revealed a range of obliquities, depending on the stellar surface temperature (e.g. \citealt{2011AJ....141...63W,2015ApJ...801....3M}).

In our simulations, we find that the obliquity distributions of the HJs and tidally disrupted planets are distinct from non-migrating systems. The obliquities of the former are broadly distributed between $\sim 0^\circ$ and $\sim 140^\circ$, with a preference for obliquities around $30^\circ$ and $60^\circ$ for HJs, and $50^\circ$ for tidal disruption systems. In contrast, the observed HJ obliquity distribution peaks around $20^\circ$ (cf. the black crosses in \F\,\ref{fig:spin}).

There is no clear peak around $130^\circ$, as found e.g. by \citet{2016MNRAS.456.3671A}, who considered high-$e$ migration in stellar binaries (cf. the black open circles in \F\,\ref{fig:spin}). This can be attributed to the fact that in our simulations, the planets were always initially prograde, and there is no expected characteristic symmetric inclination for secular evolution in multiplanet systems. Nevertheless, we still find that $\sim 0.1$ of the HJs systems have retrograde obliquities.

\section{Discussion}
\label{sect:discussion}
\subsection{Uncertainties in the secular integrations} 
\label{sect:discussion:uncertainties}
Here, we discuss uncertainties associated with the stopping condition $K_{ij}\leq 0$ and premature terminations of the integrations because the run time exceeded our maximum set value (cf. \S\,\ref{sect:pop_syn:SC}).

A comparison of the HJ fractions between runs with and without the $K_{ij}\leq 0$ stopping condition enabled in Table \ref{table:MC_grid_results_K_ij} shows that disabling the stopping condition results in different fractions. Typically, the fraction increases, which can be understood by noting that systems in which $K_{ij}\leq 0$ occurs are likely to produce high eccentricities in the inner orbit, and therefore HJs. Nonetheless, the fractions remain small, not reaching values larger than 0.007 (compared to the largest fraction of 0.004 for the simulations in Table \ref{table:MC_grid_results_K_ij} with the $K_{ij}\leq 0$ stopping condition enabled), indicating that the result of small HJ fractions is robust. Nevertheless, it remains unclear how the results would be affected if $N$-body integrations were used, which evidently do not suffer from limitations associated with small or negative $K_{ij}$. This important aspect should be investigated in future work. 

The fraction of systems in which the run time was exceeded is typically less than 0.01 (cf. Tables \ref{table:MC_grid_results} and \ref{table:MC_grid_results_K_ij}). These systems show a strong preference for large initial $a_1$, typically $a_1\sim 3.5 \, \mathrm{AU}$, and small $a_2$, typically $a_2 \sim 6.5 \, \mathrm{AU}$  (cf. \F\,\ref{fig:MC_initial_smas_es}). This implies short secular time-scales in the innermost orbit. Consequently, the number of oscillations within the time-span of 10 Gyr is very large, requiring much computation time and thus hitting our set limit of 12 CPU hours. 

Also, the number of systems in which the run time was exceeded typically increases with $\beta$. For large $\beta$, the initial eccentricities and inclinations are small, implying relatively weak secular excitation. This is also reflected by the AMD -- systems in which the run time was exceeded typically have small AMD, $\mathrm{AMD} \lesssim m_1 \sqrt{a_1}$ (cf. \F\,\ref{fig:MC_AMD_cases_MC02d}). Also, $m_1$ tends to be large, typically $m_1 \sim 4 \, M_\mathrm{J}$ (cf. \F\,\ref{fig:m_1_distributions_combined_MC02d}). 

The HJ systems, on the other hand, are associated with $a_1 \lesssim 2 \, \mathrm{AU}$ and no strong preference for $a_2$ (cf. \F\,\ref{fig:MC_initial_smas_es}), a small $\beta$, large $\mathrm{AMD}\gtrsim m_1 \sqrt{a_1}$ (cf. \F\,\ref{fig:MC_AMD_cases_MC02d}), and $m_1 \lesssim 2 \, M_\mathrm{J}$. This suggests that the HJ fractions would likely not be very different if the stopping condition (v) had not been imposed. 

In addition, we also carried out a subset of simulations with a shorter maximum CPU run time of 4 hours. We found that decreasing the maximum CPU time increases the fraction $f_\mathrm{run\,time\,exceeded}$ and decreases $f_\mathrm{no\,migration}$, whereas $f_\mathrm{HJ}$ is not strongly affected.

\subsection{HJ fraction and comparisons to other variants of high-$e$ migration}
\label{sect:discussion:comp}
In our population synthesis simulations, the highest intrinsic HJ fraction obtained was 0.023, assuming $t_\mathrm{V,1} \approx 1.4\times 10^{-3} \, \mathrm{yr}$. This corresponds to 1000 times more efficient tides compared to $t_\mathrm{V,1}\approx 1.4\,\mathrm{yr}$, which would circularize a HJ at 5 d in less than 10 Gyr \citep{2012arXiv1209.5724S}. For most other parameter combinations with non-zero fractions, the fractions are between $\sim 0.001$ and $\sim 0.01$. For the purposes of this section, we adopt a fraction $f_\mathrm{HJ,multi,sim}= 0.01$, taking into account that this fraction can be higher by a factor of at most a few if tidal dissipation in the innermost planet is extremely effective ($t_\mathrm{V,1} \lesssim 1.4 \times 10^{-2} \, \mathrm{yr}$).

Assuming a giant planet occurrence rate around MS stars of $f_\mathrm{GP} = 0.1$ and an optimistic multiplanet fraction of 1, we find a HJ fraction around MS stars of $f_\mathrm{HJ,multi} = f_\mathrm{HJ,multi,sim} f_\mathrm{GP} = 0.001$. In contrast, the observed HJ fraction is $f_\mathrm{HJ,obs}\sim 0.01$ (e.g. \citealt{2012ApJ...753..160W}), an order of magnitude larger. We emphasize that higher HJ fractions would be obtained in simulations with even smaller values of $t_\mathrm{V,1}$ (i.e. even more efficient tides), and/or larger planetary radii $R_1$. Also, we (necessarily) made assumptions about the orbital configurations (most importantly the semimajor axes, eccentricities and inclinations), which also affect the simulated HJ fractions. 

Our adopted simulated HJ fraction of $\sim0.01$ is similar or slightly lower compared to studies of high-$e$ migration in two-planet or stellar binary systems. \citet{2016MNRAS.456.3671A} find a fraction of $f_\mathrm{HJ,bin,sim}\sim 0.03$ for high-$e$ migration in stellar binaries. In \citet{2016ApJ...829..132P}, the two-planet case is considered, and $f_\mathrm{HJ,two-p,sim}=0.051$ is found for $t_\mathrm{V,1}=1.4\,\mathrm{yr}$. The high fraction in the latter paper may be due to higher assumed initial inclinations and more compact systems (smaller $a_2/a_1$) compared to our simulations. The two-planet case is also considered by \citet{2016AJ....152..174A}, who find $f_\mathrm{HJ,two-p,sim} = 0.01$ for $t_\mathrm{V,1} = 1.4 \, \mathrm{yr}$.

\subsection{Effects of disc evolution}
\label{sect:discussion:disc}
In young stellar systems (age $\lesssim 10 \, \mathrm{Myr}$), a gas disc is still present and this affects the orbital evolution of the planets (e.g. \citealt{2010ApJ...714..194M}). In the simulations of \S\,\ref{sect:ver}, the HJ formation time was typically $\lesssim 10 \, \mathrm{Myr}$ (cf. \F\,\ref{fig:nbody_ref_14_03_16_120Myr_d_tV0_HJ_formation_times}), suggesting that the effects of a dissipating gas disc should have been taken into account. We emphasize that the purpose of the simulations in \S\,\ref{sect:ver} was to test the \textsc{SecularMultiple} algorithm from a computational point of view, focusing only on the gravitational dynamics and tidal evolution. The initial conditions might not be realistic, given the large assumed initial eccentricities and inclinations (ranging between $\sim 0.4$ and $\sim 0.6$). We consider the initial conditions to be more realistic in \S\,\ref{sect:pop_syn}, in which smaller initial inclinations and eccentricities were assumed. In that section, the HJ formation times are much longer, i.e. at least $\sim 100 \,\mathrm{Myr}$ (cf. \F\,\ref{fig:end_times_MC02d}), and therefore the effects of a dissipating gas disc are likely not important.

\subsection{HJs, hot Neptunes and super-Earths from tidally downsized HJs}
\label{sect:discussion:TD_down}
Depending on the parameters, the fraction of tidally disrupted planets in our simulations can be large, typically a few times larger compared to the HJ fraction (cf. Table \ref{table:MC_grid_results}). We adopted the tidal disruption threshold from \citet{2011ApJ...732...74G}, who assumed coreless planets. In \citet{2013ApJ...762...37L}, it was found that during close encounters with their host star, Jupiter-like planets with massive cores (order 10 Earth masses) can retain part of their envelope. Consequently, the planet would not be completely tidally disrupted, but transformed into a low-mass HJ, a hot Neptune or a super-Earth, depending on the amount of mass lost.

When related to our simulations, this suggests that there could be a significant contribution to low-mass HJs, hot Neptunes or (short-period) super-Earths driven by secular evolution in multiplanet systems. This does require that the original planet was not too massive ($\lesssim 2 \, M_\mathrm{J}$, cf. the yellow lines in \F\,\ref{fig:m_1_distributions_combined_MC02d}), and the typical formation time of the tidally downsized planet is expected to be much shorter compared to that of HJs formed through tidal evolution (cf. \F\,\ref{fig:end_times_MC02d}).

\section{Conclusions}
\label{sect:conclusions}
We have studied the orbital migration of Jupiter-like planets induced by secular interactions in multiplanet systems (three to five planets), resulting in HJs. In this variant of high-$e$ migration, the eccentricity of the orbit of the innermost planet is excited to high values on secular time-scales (order Gyr). Combined with tidal dissipation, which is highly effective for high eccentricities, this can produce a Jupiter-like planet in a tight orbit. Our conclusions are as follows.

\medskip \noindent 1. For a set of three-planet systems we have shown that the secular code \textsc{SecularMultiple} \citep{2016MNRAS.459.2827H} produces results that are statistically consistent with those of more accurate direct $N$-body integrations (\S\,\ref{sect:ver}).  

\medskip \noindent 2. We carried out a population synthesis study of multiplanet systems with \textsc{SecularMultiple}, taking into account tidal dissipation in the innermost planet and the central star (\S\,\ref{sect:pop_syn}). We found HJ fractions of at most 0.023, assuming $t_\mathrm{V,1} \approx 1.4\times 10^{-3} \, \mathrm{yr}$. This corresponds to 1000 times more efficient tides compared to $t_\mathrm{V,1}\approx 1.4\,\mathrm{yr}$, for which a HJ at 5 d would circularize in less than 10 Gyr \citep{2012arXiv1209.5724S}. For relatively weak tidal dissipation ($t_\mathrm{V,1} \gtrsim 1.4 \times 10^{-2} \, \mathrm{yr}$), we found no HJs. Larger fractions would be obtained for even lower values of the innermost planet viscous time-scale (stronger tides), i.e. $t_\mathrm{V,1} \lesssim 1.4\times 10^{-3} \, \mathrm{yr}$. The HJ fractions are similar or lower compared to other variants of high-$e$ migration, but this comparison depends strongly on the system parameters, such as the initial semimajor axes, eccentricities, inclinations, the viscous time-scales and the radius of the innermost planet.

\medskip \noindent 3. In the population synthesis, we found that the fraction of systems in which the innermost planet is tidally disrupted is typically a few times larger compared to the HJ fraction. The large proportion of tidally disrupted planets to HJs can be explained qualitatively by noting that the eccentricity of the innermost orbit can be violently excited in multiplanet systems, implying that the planet is rapidly tidally disrupted before tidal dissipation is able to shrink and circularize the orbit. The large fraction of tidal disruptions in some of our simulations suggests that tidal disruptions in multiplanet systems, even if not extremely compact, could be common. For non-compact planetary systems, this suggests a possible difference in metallicity between stars with two or fewer planets, compared to stars with three or more planets. 

\medskip \noindent 4. The orbital period distribution of the HJs in our simulations is strongly peaked around $\sim 5 \, \mathrm{d}$, which coincides with the peak in the observed orbital period distribution of gas giant planets. The location of the peak is affected by the assumed tidal dissipation efficiency and the planetary radius. In our simulations, HJs with the longest periods correspond to an inflated planet with radius $R_1 = 1.5\,R_\mathrm{J}$. No significant number of planets was found in the simulations with orbital periods in the `period valley' between 10 and 100 d, whereas observations show a significant population of planets in this regime, i.e. WJs. It is unlikely that WJs are produced through secular evolution in multiplanet systems, unless tidal dissipation is extremely efficient. Other high-$e$ migration scenarios also fail to produce WJs in the observed proportion \citep{2016ApJ...829..132P,2016AJ....152..174A}. Alternative candidates for the origin of WJs are {\it in situ} formation or disc migration. 

\medskip \noindent 5. Our simulated HJs and tidally disrupted planets are preferentially not massive, i.e. $m_1\lesssim 2 \, M_\mathrm{J}$, with a median value of $\approx 1 \, M_\mathrm{J}$ (cf. \S\,\ref{sect:pop_syn:results:mass}), which is similar to observations. The stellar obliquity distribution is fairly uniform between $\sim 0^\circ$ and $\sim 140^\circ$ with some preference for obliquities around $30^\circ$ and $60^\circ$. There is no clear peak at $\sim130^\circ$, as opposed to high-$e$ migration in stellar binary or two-planet systems (cf. \S\,\ref{sect:pop_syn:results:spin}). Approximately 0.1 of the HJs have retrograde obliquities.

\medskip \noindent 6. Another characteristic of HJs formed in our simulations relevant for observations is the late formation time of up to $\sim 10\,\mathrm{Gyr}$ (cf. \S\,\ref{sect:pop_syn:results:times}). This is in stark contrast to disc migration, for which formation is expected to occur within the first few Myr. Also, this characteristic can potentially distinguish between other variants of high-$e$ migration, which typically predict shorter formation times.

\section*{Acknowledgements}
We thank the anonymous referee for providing helpful and constructive comments. This work was supported by the Netherlands Research Council NWO (grants \#639.073.803 [VICI],  \#614.061.608 [AMUSE] and \#612.071.305 [LGM]) and the Netherlands Research School for Astronomy (NOVA). We acknowledge the use of the astric computer cluster of the national Israeli astrophysics I-CORE centre. HBP acknowledges support from European union career integration grant `GRAND', the Minerva centre for life under extreme planetary conditions and the Israel science foundation excellence centre I-CORE grant 1829. YL acknowledges NSF grants AST-1109776 and AST-1352369 and NASA grant NNX14AD21G.

\bibliographystyle{mnras}
\bibliography{literature}

\label{lastpage}
\end{document}